\documentstyle[preprint,eqsecnum,aps]{revtex}

\def\simlt{\stackrel{<}{{}_\sim}}
\def\simgt{\stackrel{>}{{}_\sim}}

\def\slashchar#1{\setbox0=\hbox{$#1$}           
   \dimen0=\wd0                                 
   \setbox1=\hbox{/} \dimen1=\wd1               
   \ifdim\dimen0>\dimen1                        
      \rlap{\hbox to \dimen0{\hfil/\hfil}}      
      #1                                        
   \else                                        
      \rlap{\hbox to \dimen1{\hfil$#1$\hfil}}   
      /                                         
   \fi}                                         %
\def\tightenlines{\def\baselinestretch{1.3}\small\normalsize}
\tightenlines
\begin{document}
\input psfig.sty
\preprint{\vbox{\hfill ANL--HEP--PR--98--54 \\\vbox{\hfill CERN--TH/98--262} \\\vbox{\hfill FERMILAB--PUB--98/250--T}}}
\title{MSSM Higgs Boson Phenomenology at the Tevatron Collider}
\author{M. Carena\cite{mcemail}\\
Fermi National Accelerator Laboratory,
Batavia, IL 60510 USA}
\author{
S. Mrenna\cite{byline}\\
Argonne National Laboratory,
Argonne, IL 60439  USA}
\author{C.E.M. Wagner\cite{cwemail}\\
CERN, TH Division,
CH--1211 Geneva 23, Switzerland}
\date{\today}
\maketitle

\begin{abstract}
The Higgs sector of the minimal supersymmetric standard model (MSSM)
consists of five physical Higgs bosons, which offer a variety of channels
for their experimental search. 
The present study aims to further our understanding
of the Tevatron reach for MSSM Higgs bosons, addressing relevant 
theoretical issues related to the SUSY parameter space,
with special emphasis
on the radiative
corrections to the down--quark and lepton couplings to the Higgs bosons
for large $\tan\beta$.
We performed a computation of the signal and backgrounds for the production 
processes $W\phi$ and $b \bar{b} \phi$ at the upgraded Tevatron, with $\phi$  
being the neutral MSSM Higgs bosons.
Detailed experimental information and further higher order calculations 
are demanded to confirm/refine these predictions.

\end{abstract}

\pacs{ }

\narrowtext

\section{Introduction}
\label{sec:intro}

The precision electroweak measurements performed at  LEP, SLD and
the Tevatron are consistent with the predictions of the
standard model containing a light Higgs boson, with mass
of the order of the $Z$ boson mass. 
The searches for such a Higgs particle continue
at the LEP and the Tevatron colliders. 
The searches at LEP2 ($\sqrt{s} \simlt 200$ GeV) 
are constrained by the collider energy, and a Higgs boson
with standard model--like properties can be found only if its
mass is below 105 GeV~\cite{CARZER}. 

The potential for discovering a light Higgs boson at the
Tevatron collider when it is produced in association with a $W$ or $Z$ gauge
boson has been discussed in several studies
\cite{higgs1,tevus,tev2000,kuhlmann,yao}.
Although the kinematic reach of the Tevatron
collider is much greater than for LEP2, the backgrounds to
Higgs boson  searches at hadron colliders are much larger than in $e^+ e^-$ machines.
For this reason, large integrated luminosity is essential to establish
a signal at the Tevatron. 
Within the standard model, the general conclusion is that Run II,
with a total integrated luminosity of about 2 fb$^{-1}$ per detector, will be 
unable to extend the Higgs boson mass reach of LEP2. 
The main questions
are:  what is the theoretical motivation for a Higgs boson with a mass 
slightly above the LEP2 reach, and
what is the necessary upgrade in luminosity to cover that region?

We address the theoretical motivation by appealing to 
the minimal supersymmetric extension of the standard model (MSSM).
The MSSM has
the remarkable property that, for a sufficiently heavy supersymmetric
spectrum, it fits to the precision electroweak observables
as well as the standard model\cite{ewfit}. Moreover, the lightest CP--even
Higgs boson mass $m_h$ is constrained to satisfy $m_h \simlt 130$ 
GeV\cite{CAESQUWA,HHH}. The 
Higgs sector of this model consists of two Higgs doublets, with
two CP--even Higgs bosons, $h$ and $H$, one CP--odd Higgs boson, $A$, and 
one charged Higgs boson, $H^{\pm}$. 
This richer spectrum allows for different production and decay
processes at LEP and the Tevatron colliders
than in the standard model.

In the supersymmetric
limit, the neutral components of the two 
Higgs boson doublets $H_1$ and $H_2$ couple 
to  down-- and up--type
quarks, respectively. Lepton fields couple only to the $H_1$ 
Higgs boson.  The MSSM, tree--level Yukawa couplings of the down
quarks, leptons and up quarks are
related to their respective running masses by
\begin{eqnarray}
h_d & \simeq & \frac{m_d}{v \cos\beta}, \;\;\;\;\;\;\;\;\;\;\;
h_l  \simeq  \frac{m_l}{v \cos\beta},   \;\;\;\;\;\;\;\;\;\;\;
h_u  \simeq  \frac{m_u}{v \sin\beta},
\label{tree}
\end{eqnarray}
where $\tan\beta = v_2/v_1$ is the ratio of the vacuum expectation
values of the two Higgs doublets, and $v = \sqrt{v_1^2 + v_2^2}$=174 GeV. 
In the standard model, 
only the top quark Yukawa coupling $h_t$ 
is of order one at the weak scale.
In the MSSM, instead, the bottom and $\tau$ Yukawa couplings,
$h_b$ and $h_\tau$,
can become of the same order as $h_t$, if $\tan\beta$
is sufficiently large.   This can have important phenomenological
consequences.

Quite generally, the two CP--even Higgs boson eigenstates are
a mixture of the real, neutral $H_1$ and $H_2$ components,
\begin{eqnarray}
\left( \begin{array}{c} h \\ H \end{array} \right)= 
\left( \begin{array}{cc} -\sin\alpha & \cos\alpha \\
\cos\alpha & \sin\alpha \end{array} \right)
\left( \begin{array}{c} H_1^0 \\ H_2^0 \end{array} \right),
\label{mixings}
\end{eqnarray}
and the lightest CP--even Higgs boson couples to down quarks (leptons),
and up quarks by its standard model values
times  $-\sin\alpha/\cos\beta$ and $\cos\alpha/\sin\beta$, respectively.
The couplings to the heavier CP--even Higgs boson are given by the standard 
model values
times $\cos\alpha/\cos\beta$ and $\sin\alpha/\sin\beta$, respectively.
Analogously, the coupling of the CP--odd Higgs boson to down quarks (leptons)
and up quarks is
given by the standard model coupling times $\tan\beta$
and $1/\tan\beta$, respectively. Moreover, the
lightest (heaviest) CP--even Higgs boson has  $ZZh$ and
$WWh$ ($ZZH$ and $WWH$) couplings
which are given by the Standard Model value times $\sin(\beta - \alpha)$
($\cos(\beta - \alpha)$), while it can be produced in association with
a CP--odd Higgs boson with a $ZhA$ ($ZHA$) coupling which is proportional
to $\cos(\beta - \alpha)$ ($\sin(\beta - \alpha)$). 

For sufficiently large values of the CP--odd Higgs boson mass $m_A$, 
the effective theory at low energies contains only one Higgs doublet, with 
standard model--like properties, in the combination
\begin{equation}
h \simeq \phi^{SM} = H_1^0 \cos\beta + H_2^0 \sin\beta; \;\;\;\;\;\;\;
<\phi^{SM}> = v\; ,
\label{vev}
\end{equation}
where $\sin \alpha \simeq - \cos \beta$, $\cos \alpha \simeq \sin \beta$
and hence $\sin^2(\beta - \alpha) = 1$.
In this limit, if all supersymmetric particles are heavy, 
all phenomenological conclusions drawn
for a SM Higgs boson are robust when extended to the the lightest CP--even
Higgs boson of the 
MSSM, which, as mentioned before, is at the same time constrained to 
have a mass below
about 130 GeV.
For large $\tan\beta$,
one of the CP--even neutral Higgs bosons tends to be degenerate in mass with
the CP--odd Higgs boson and couples strongly to the bottom quark
and tau lepton. The other CP--even Higgs boson has standard
model--like couplings to the gauge bosons, 
while its coupling to the down quarks and leptons may be 
highly non--standard. 
If $m_A$ is large, then $h$ is the Higgs boson
with SM--like properties as described 
above.
If $m_A$ is small, then $H$ is the 
one with SM--like couplings to the gauge bosons. 
In the following, the symbol $\phi$ denotes a generic Higgs boson.

In this article, we analyze the discovery potential of the
Tevatron collider for MSSM Higgs bosons 
in different production channels.
Section 2 discusses the signals from
Higgs boson production in association with
weak gauge bosons and their dependence on the MSSM parameter space.
Section 3 contains details of Yukawa coupling effects for large $\tan\beta$.
Section 4 deals with the phenomenological implications of these
effects for signals in the $W\phi$ and $b\bar b\phi$ production
channels.  In section 5, we consider the correlation between the 
bottom mass corrections and the supersymmetric contributions to the
branching ratio
$B(b \rightarrow s \gamma)$.  Finally, section 6 is reserved
for our conclusions.

\section{Signals from $W\phi$ and $Z\phi$ Production}

The production of $W\phi$, followed by the decays
$W(\to e\nu_e)$ or $W(\to \mu\nu_\mu)$ and $\phi(\to b\bar b)$,
is the gold--plated search mode for the standard model Higgs
boson $\phi=\phi^{SM}$ at the Tevatron collider,    
while LEP2 is sensitive to
the $Z\phi^{SM}$ process.
The production of $Z\phi$ may also be useful at the
Tevatron, depending
on the efficiency for triggering on missing $E_T$ 
($\slashchar{E}_T$) and the $b\bar b$ mass resolution.  
Additionally, the
all--hadronic decays of $W\phi+Z\phi$ may extend the reach,
and the Higgs boson decay $\phi(\to\tau^+\tau^-)$ may be
observable.  However, there are
several, unresolved experimental issues concerning these
channels that require detailed study by the experimental
collaborations.  
For this reason, at present we shall only consider
the $W\phi$ channel at the Tevatron.

To quantify the experimental reach in a model independent way, 
it is useful to consider the function 
\begin{equation}
R= \frac{\sigma(p \bar p \to W \phi)}{\sigma(p \bar p \to W\phi^{SM})} 
 \frac{B(\phi\to b\bar b)}{B(\phi^{SM}\to b\bar b)},
\label{eq:R}
\end{equation}
where $\sigma$ denotes a 
production cross section,  $B$ denotes a branching ratio and $\phi^{SM}$ 
represents a standard model Higgs boson.
In the MSSM, the ratio of cross sections is just given by $
\sin^2(\beta-\alpha)$ or $\cos^2(\beta-\alpha)$ depending on $\phi$ being 
the lightest or heaviest CP--even Higgs, respectively,
\footnote{Observe that 
$\sin(\beta - \alpha)$ ($\cos(\beta - \alpha)$)
denotes the component of the
lightest (heaviest)
CP--even Higgs boson in the combination which acquires
a vacuum expectation value, Eq. (\ref{vev}).}
while the ratio of branching ratios has a more complicated behavior.
It is important to notice that,
barring the possibility of large next--to--leading--order (NLO), SUSY 
corrections to the $W W \phi$ vertex,
there is no enhancement of the 
production cross section in the MSSM over the standard model.
On the other hand, the branching ratio to
$b\bar b$ and $\tau^+\tau^-$ final states are affected by the  factors  
$-\sin \alpha/\cos \beta$ for $h$ and $\cos \alpha/\cos \beta$ for $H$ over 
the SM Higgs boson 
couplings to down quarks and leptons. These factors can produce an
increase or decrease of the MSSM coupling of the Higgs boson to
bottom quarks, depending 
on the value of the CP--odd mass,  $\tan \beta$ and the top and bottom
squark mass parameters. 
In this study, the Higgs boson properties are calculated
  using the program {\tt Hdecay}\cite{hdecay}.

As mentioned above, for large $m_A$,
the low--energy, effective theory contains only
one Higgs boson with SM--like properties. 
The Yukawa couplings tend to the
standard model values, and   $\sin^2(\beta-\alpha)\simeq 1$.  
As long
as no new decay modes are open, $h$ has the same  properties as $\phi^{SM}$,
and a discovery or exclusion limit for a standard model Higgs boson applies
equally well to $h$, and $R=1$.  
If Higgs boson decays to sparticles become 
important, then it is quite likely that additional Higgs boson production
modes exist and enhance the potential signal, rather than decrease it.
For example, if the sparticle spectrum is of the order of $m_h$,
then processes like $p\bar p\to N_2(\to N_1 h) C_1(\to N_1 W)+X$ can occur.
In our analysis, we shall always consider the limit of heavy sparticle
masses, where such supersymmetric contributions to the Higgs
production and decay processes are negligible.

In the large $m_A$ limit,
the  renormalization--group improved result for
the lightest Higgs boson mass, including two--loop 
leading--log effects~\cite{CAESQUWA,HHH,HEHO,CAESQURI},
has the approximate analytic form~\cite{CAESQUWA}:
\begin{eqnarray}
m_h^2 \equiv m_{\phi^{SM}}^2 & = & M_Z^2\cos^2(2\beta)
\left(1-{3\over 8\pi^2}{m_t^2\over v^2}t\right) \nonumber\\
& + & {3\over 4\pi^2}{m_t^4\over v^2}\left[ {1\over 2}\widetilde{X}_t + t +
{1\over 16\pi^2}\left({3\over 2}{m_t^2\over v^2}-32\pi\alpha_3\right)
\left(\widetilde{X}_t t 
+ t^2\right)\right], \nonumber\\
\widetilde{X}_t & = & 
{2\tilde{a}^2}
\left(1-{\tilde{a}^2\over 12}\right), \;\;\;\;\;\;\;\;
\tilde{a} = \bar{A}_t-\bar{\mu}/\tan\beta,
\label{higgsm2l}
\end{eqnarray}
where $\bar{\mu} = \mu/M_S$ and 
$\bar{A}_t = A_t/M_S$,
with $M_S^2 = (m_{\tilde t_2}^2 + m_{\tilde t_1}^2)/2$, and
$t=\log(M_S^2/m_t^2)$. 
The above formula is based on an approximation in which 
the right--handed and left--handed stop supersymmetry breaking
parameters are assumed to be close to each other, and hence
the stop mass splitting is 
induced by the mixing parameter $m_t \tilde{a}\times M_S$.
Moreover, 
this expression is based on an expansion in
powers of $m_t \tilde{a}/M_S$  and 
is valid only if
$\frac{|m_{\tilde t_1}^2 - m_{\tilde t_2}^2|}
{m_{\tilde t_2}^2 + m_{\tilde t_1}^2}
< 0.5$, where $m_{\tilde t_1}$ and $m_{\tilde t_2}$ are
the lightest and heaviest stop mass eigenstates.
This simplified
expression  is very useful for 
understanding  the results of this work,
although  we go beyond this
approximation~\cite{CAESQUWA} in our full analysis. Finally, in
the above, we have ignored corrections induced by the sbottom
sector, which, as we shall discuss below, may become relevant for
very large values of $\tan\beta$.

The value of $m_h$ in Eq. (2.2) is maximized for 
large values of $\tan\beta$ and  $M_S$
and  $\tilde{a}^2 = 6$. Due to the dependence of the lightest
CP--even Higgs mass on $\tan\beta$, LEP will be able to probe the 
low $\tan\beta$ region of the MSSM.
Indeed, recent analyses suggest that even the present, relatively
low bounds on a standard model--like Higgs boson
from LEP2 have strong implications for the minimal supergravity
model~\cite{Ellis}. Moreover, it has been shown that, in the large
$m_A$ region, LEP2 will probe values of $\tan\beta \simlt 2$ for
arbitrary values of the stop masses and mixing angles~\cite{CACHPOWA}. 
Since, in general, the lower bound on $\tan\beta$ is expected to be obtained
for large values of $m_A$, the absence of a Higgs signal in the
$ZZh$ channel at LEP2 will provide a strong motivation for models with
moderate or large values of $\tan\beta$.

For lower values of the CP--odd Higgs mass, 
$\sin^2(\beta-\alpha)$ can take any value
between 0 and 1, and it is a model--dependent question whether $h$ or
$H$ produces a viable signal in the $W\phi$ production channel. For moderate
or large values of $\tan\beta \simgt 5$, it is easy to identify the
main properties of the CP--even Higgs sector.   
More specifically, three cases may occur:

~\\
a) If $m_A < m_{\phi^{SM}}$, then
$\sin\alpha \simeq - 1$, 
$\cos\alpha \simeq 
{\cal O}(1/\tan\beta)$ and $\cos(\beta - \alpha) \simeq 1$.
In this case, the {\it heaviest} CP--even Higgs boson has a production rate which
is similar to the standard model case. The branching ratio of the decay
into bottom quarks and $\tau$ leptons, however, can become highly
non--standard, since $\cos\alpha$ and $ \cos\beta$,  
may differ by a factor of order one.

~\\
b) If $m_A > m_{\phi^{SM}}$, then the {\it lightest} CP--even Higgs boson 
has a production rate similar to the standard model case. 
For $m_A \gg m_{\phi^{SM}}$, the
branching ratio of the decay of this Higgs boson into down quarks
is standard model--like.   However, when  
$m_A$ becomes close to $m_{\phi^{SM}}$, there 
can be important
differences in the branching ratios with respect to the SM ones.

~\\
c) If  $m_A \simeq m_{\phi^{SM}}$, then 
$\sin^2(\beta - \alpha) \simeq \cos^2(\beta - \alpha) \simeq 
{\cal O}( 0.5 )$, 
and the couplings of both neutral CP--even Higgs bosons to bottom quarks
tend to be highly non--standard.

To better understand the behavior of the Higgs boson branching
ratios in these different cases, we analyze the Higgs boson mass matrix.
Assuming the approximate conservation
of CP in the Higgs sector, the CP--even Higgs masses may be determined
by diagonalizing the 2 $\times$ 2 symmetric mass matrix ${\cal M}^2$.
After including the dominant one--loop corrections induced by the 
stop and sbottom sectors, together with the two--loop, 
leading--logarithm effects, the elements of ${\cal M}^2$ 
are \cite{CAESQUWA}
\begin{eqnarray}
{\cal M}^2_{11} & \simeq & m_A^2 \sin^2\beta + M_Z^2 \cos^2\beta
\nonumber\\
& - &\frac{h_t^4 v^2}{16 \pi^2 } \bar{\mu}^2 \sin^2\beta\tilde{a}^2
\left[ 1 + \frac{t}{16 \pi^2}\left( 6 h_t^2 - 2 h_b^2 - 16 g_3^2
\right) \right]
+ {\cal O}(h_t^2 M_Z^2)
\nonumber\\
{\cal M}^2_{22} & \simeq & m_A^2 \cos^2\beta + M_Z^2 \sin^2\beta
\left(1 - \frac{3}{8 \pi^2} h_t^2 t \right)
\nonumber\\
& + &\frac{h_t^4 v^2}{16 \pi^2} 12 \sin^2\beta \left\{ 
t \left[ 1 + \frac{t}{16 \pi^2} \left( 1.5 h_t^2 + 0.5 h_b^2 - 8 g_3^2 \right)
\right]
\right.
\nonumber\\
& + & \left. 
\bar{A}_t\tilde{a}\left(1  - {\bar{A}_t\tilde{a} \over 12}\right)
\left[ 1 + \frac{t}{16 \pi^2} \left( 3 h_t^2 +  h_b^2 - 16 g_3^2 \right)
\right] \right\}
\nonumber\\
& - &   \frac{v^2 h_b^4}{16 \pi^2} \sin^2\beta \bar{\mu}^4
\left[ 1 + \frac{t}{16 \pi^2} \left( 9 h_b^2 - 5 h_t^2 - 16 g_3^2 \right)
\right]  
+ {\cal O}(h_t^2 M_Z^2)
\nonumber\\
{\cal M}^2_{12} & \simeq & -\left[m_A^2 + M_Z^2
- \frac{h_t^4 v^2}{8 \pi^2} \left(3 \bar{\mu}^2 - \bar{\mu}^2 \bar{A}_t^2 
\right)\right] 
\sin\beta \cos\beta
\nonumber\\
& + & \left[
\frac{h_t^4 v^2}{16\pi^2}\sin^2\beta \bar\mu \tilde{a}\left[\bar{A}_t
\tilde{a} - 6 \right] + \frac{3 h_t^2 M_Z^2}{32 \pi^2} 
\bar{\mu} \tilde{a} \right] 
\left[ 1 + \frac{t}{16 \pi^2} \left( 4.5 h_t^2 - 0.5 h_b^2 - 16 g_3^2 \right)
\right],
\label{matel}
\end{eqnarray}
where $g_3$ is the QCD running coupling constant. In the
above, we have assumed, for simplicity, that 
$A_b \simeq 0$ (an assumption we shall always make in the following
analysis)
and retained only the leading terms in powers of $h_b$ and $\tan\beta$.
We have also included the small, ${\cal O}(h_t^2 M_Z^2)$ correction
to ${\cal M}_{12}^2$ explicitly because it plays a relevant role in
our analysis.
The above expressions hold only in the limit of small splittings
between the running stop masses. Moreover, the
condition $2m_t\max(|A_t|,|\mu|)<M^2_S$ must be fulfilled.
Similar conditions should be fulfilled in the sbottom sector.
The leading, two--loop, logarithmic corrections to the squared Higgs
mass matrix elements included above can
be as large as $10-20\%$ when supersymmetric particles are heavy,
and are very relevant in determining
the Higgs boson mass eigenvalues
and mixing angles. Observe that Eq.~(\ref{higgsm2l}) may be easily 
obtained from the above expression, by computing the determinant of
the Higgs boson mass matrix and 
setting the heavy CP--even Higgs boson mass approximately 
equal to $m_A$.\footnote{There is, however, a slight
discrepancy in the subdominant, Yukawa--dependent, two--loop, 
leading--logarithmic
corrections, which is due to the fact that the expressions written above
are only strictly valid for values of the CP--odd mass of the order
of the weak scale.}

The mixing
angle $\alpha$ can be determined from the expression
\begin{eqnarray}
\sin\alpha\cos\alpha = {{\cal M}_{12}^2 \over 
\sqrt{({\rm Tr}{\cal M}^2)^2-4\det{\cal M}^2}}.
\label{sin2alpha}
\end{eqnarray}
In the limit that ${\cal M}_{12}\to 0$, either $\sin\alpha$ or 
$\cos\alpha\to 0$.  For moderate or large values of
$\tan\beta$, 
if case a)  is realized and 
$\cos\alpha \simeq 0$, the coupling of the 
standard model--like Higgs boson\footnote{From now on, the
term ``standard model--like Higgs boson'' refers to a Higgs boson with
standard model couplings to the gauge bosons with no implication about
its couplings to fermions.} $H$ to $b\bar b$ and
$\tau^+\tau^-$ is diminished, and decays to
$gg$, $\gamma\gamma$, $c\bar c$, and $W^{(*)}W^{(*)}$ can be
greatly enhanced over standard model expectations~\cite{Wells}.
The same can happen for $h$
in case b) when $\sin\alpha \simeq 0$.
For moderate or large values of 
$\tan\beta$, the vanishing of ${\cal M}_{12}^2$ leads
to the approximate numerical relation:
\begin{equation}
\left[{m_A^2\over m_Z^2} - \frac{1}{2 \pi^2} \left( 3 \bar{\mu}^2 -
\bar{\mu}^2 \bar{A}_t^2 \right) +
1 \right] \simeq {\tan\beta\over 100} \left[
\bar\mu \tilde{a} (2 \tilde{a} \bar{A}_t - 11)\right] \left[ 1 -
\frac{15}{16 \pi^2}  t \right],
\label{suppression}
\end{equation}
where we have neglected the bottom Yukawa coupling effects and 
replaced $h_t$ and $g_3$ 
and the weak gauge couplings by their approximate numerical values at the
weak scale.
For low values of $m_A$, or large values of the
mixing parameters, a cancellation can easily 
take place for large values
of $\tan\beta$. For instance, 
if $M_S \simeq 1$ TeV,
$\bar\mu=-\tilde{a}=1$, and $m_A \simeq 80$ GeV, a cancellation can
take place for $\tan\beta \simeq 28$, with the spectrum
$m_A \simeq m_h$ and $m_H \simeq 117$ GeV. The heaviest
CP--even Higgs boson has standard model--like couplings
to the gauge bosons 
($\cos^2(\beta-\alpha)\simeq 1$), but the branching ratios for
decays into $W^{\pm}$ bosons, gluons and charm quarks are 
enhanced with respect to the SM case:
$B_W=.34, B_g=.27$ and $B_c=.11$. For the same value
of $m_A$, but larger values of the stop mixing
parameters,
$\bar\mu=\tilde{a}=\sqrt{7}$ (at the edge of the region of validity
of the above approximation), 
an approximate cancellation of the 
tree--level bottom and $\tau$ lepton couplings
is achieved for 
$\tan\beta=20$, for which $m_H\simeq 124$ GeV, with branching
ratios $B_W=.57, B_g=.21,$ and $B_c=.08$.

An interesting point is that, for large values of $\tan\beta$ and 
values of the stop mixing parameters which maximize ${\cal M}_{22}^2$,
($\tilde{a} = \sqrt{6}$), the dominant, $m_t^4$--dependent corrections to the 
off--diagonal elements of the Higgs boson matrix vanish. In this case, the corrections
to ${\cal M}_{12}^2$ 
are dominated by the $m_t^2 M_Z^2$ dependent terms
(see Eq. (\ref{matel})), which, for $M_S \simeq 1$ TeV, 
cannot be large enough to induce an approximate cancellation of the
off--diagonal terms for $|\bar\mu| \leq 1$ in the region of $m_A \simlt 400$ 
GeV and
$\tan\beta \simlt 50$ considered in this article. However, even for 
$|\bar{\mu}| = 1$, the impact of the radiative corrections to the
off--diagonal elements of the Higgs boson mass matrix may be very relevant for
low values of $m_A$, and
we expect large variations of the
branching ratio of the decay of the heaviest CP--even Higgs boson
into bottom quarks with respect to the choice of the sign of $\mu$
in this region of parameters.  

Moreover, away from
$|\tilde{a}| = \sqrt{6}$, the $m_t^4$--dependent radiative corrections
to ${\cal M}_{12}^2$ depend strongly on the sign of
$\bar\mu \times \bar A_t$ ($\bar A_t \simeq \tilde{a}$ for large $\tan\beta$ 
and moderate $\mu$)
and on the value of $|\bar{A}_t|$.
For the same value of $\tilde{a}$, 
a change in the
sign of $\mu$ can lead to observable variations in the branching
ratio for the Higgs boson decay into bottom quarks. If $|\bar{A}_t| \simlt
\sqrt{11/2}$, the absolute value of the off--diagonal matrix
element, and hence, the coupling of bottom
quarks to the standard model--like Higgs boson tends to be suppressed (enhanced)
for values of 
$\bar{\mu} \times \bar{A}_t < 0$  ($\bar{\mu} \times \bar{A}_t > 0$).  
For larger values of $|\bar{A}_t|$, instead, the suppression (enhancement)
occurs for the opposite sign of $\bar{\mu} \times \bar{A}_t$.
Finally, it is important to stress that,
in the large  $\tan \beta$ regime, extra corrections to the Yukawa 
couplings may be important depending on the MSSM spectrum, and 
we shall come back to this topic later in Section 3.

\subsection{$W(\to\ell\nu)\phi^{SM}(\to b\bar b)$ signal and backgrounds}

The signal $W(\to\ell\nu)\phi^{SM}(\to b\bar b)$
contains two real $b$--jets, which can be used
to distinguish it from many potential backgrounds.
A remarkable increase in the double tag
efficiency over the Run I estimate is expected to be 
accomplished by loosening the requirements for
the second tag\cite{kuhlmann}.  This is possible since
the first $b$--tag already significantly reduces the
fake $b$ background. In the numerical  analysis we assume a
double tagging efficiency for Run II of 
$\epsilon_{b\bar b}=.5^2\times 1.8=.45$.

Given the expected high $b$--tagging efficiency and the low mistagging
rate, the most important backgrounds are those with 
a real charged lepton $\ell^\pm$ and
two real $b$ taggable jets.  The backgrounds
considered are $W^\pm g^*(\to b\bar b),
W^\pm Z^0(\to b\bar b), 
t(\to bW^+)\bar t(\to \bar b W^-), {W^+}^*\to t(\to b W^+) \bar b,$
and $qg\to q^{'}t(\to b W^+) \bar b$,
and hermitian conjugate (h.c.) processes where appropriate.
The $W^\pm$ boson decays leptonically, $W^\pm(\to\ell^\pm\nu)$,
where $\ell^\pm = e^\pm$ or $\mu^\pm$, except in the $t\bar t$
background, where the second $W^\pm$ boson can decay hadronically or
to a $\tau$.

Events are required to have one lepton with
$p_T^{(\ell)}>$ 20 GeV and $|\eta^{(\ell)}|<2$.
The lepton must be isolated from jets with a separation
$R_{\phi} = \sqrt{\Delta\phi^2+\Delta\eta^2} > .4$, where $\Delta\phi$
and $\Delta\eta$ are the difference in azimuthal angle and 
pseudorapidity between the lepton and jets.
Jets which can be heavy flavor tagged must
have $E_T^{(b)}>15$ GeV and $|\eta^{(b)}|<2$.  Additional jets
are resolved if
$E_T^{(j)}>15$ GeV and $|\eta^{(j)}|<2.5$, and leptons if
$p_T^{(\ell)} > 10$ GeV and $|\eta^{(\ell)}|<2$.
Jets must be separated from each other by $R_{\phi} >.7$.
A Higgs boson signal is defined as an excess above backgrounds in
the invariant mass distribution of the $b\bar b$ pair, $M_{b\bar b}$.
A mass resolution
of about $\sigma_M = .08M_{b\bar b}$ is assumed.
For $M_h=100$ GeV, this means a window
$84< M_{b\bar b} < 116$ GeV.

To further enhance the signal over the background, we apply additional
cuts.
The angle between the $W^\pm$ and $h$ in the $W^\pm h$ center
of mass system, $\theta_h$, can be exploited\cite{costheta}.
The dominant background from $W^\pm g^*(\to b\bar b)$ tends
to peak at $\cos\theta_h=\pm 1$, and a cut of $|\cos\theta_h|<.8$ is
optimal.
Top quark pair production events, 
which are a large potential background, produce
an additional $W^\pm$ boson, which can decay hadronically or
leptonically.  The extra decay products from this $W$ decay can
be used to veto such events.
Vetoing events with at least one jet with $E_T > 30$ GeV, a
pair of jets separately having $E_T > 15$ GeV, or extra leptons with
$p_T > 10$ GeV successfully limits this background.

The estimate of the signal and background used in this analysis
are based on
a parton level calculation \cite{persii},
and the final state partons are interfaced to a detector simulation
to account for finite detector resolution in measuring energies
and angles\cite{d0note}.
The NLO QCD corrections to those processes
which are order $\alpha_s^0$ at tree level are large.
The order $\alpha_s$ correction to $W^\pm Z^0$ production at 
Tevatron energies is about 1.3\cite{ohnemus}.
A similar enhancement occurs for the signal process
$W^\pm h$, which has the same initial state, and
the rate for the process $u\bar d\to {W^+}^*\to t\bar b$,
with both initial and final state corrections,
is enhanced by a factor of 1.7 \cite{wtbus} over the lowest result using
CTEQ3L \cite{cteq3} structure functions.
The single top production processes $qb\to q't$
increases by a factor of about 1.3\cite{bordes}.
Results are normalized to these numbers.
In addition, $t\bar t$ and $W^\pm g^*(\to b\bar b)$ are evaluated at the 
scale $m_t$, which gives good agreement with the present data.
However, higher--order calculations of the $W^\pm g^*(\to b\bar b)$
production rate and kinematics are necessary to confirm our understanding
of the standard model backgrounds~\cite{rkellis}.

The results of this analysis will be used below.
These results are in good agreement
with other studies\cite{tev2000,kuhlmann} after accounting
for different assumptions for the mass resolutions $\sigma_M$
(see also Ref.~\cite{boos}).   

\subsection{Results on $Wh$ and $WH$}

The signal and background estimates from the analysis described above can 
be used to estimate the discovery or exclusion reach of the Tevatron.
For a fixed integrated luminosity and a Higgs boson mass, 
we can determine which values
of $R$ would lead to a discovery with a 5$\sigma$ significance 
($\sigma=S/\sqrt{B}$) or a 95\% C.L. ($1.96\sigma$) exclusion.
If the $R$--contour lies below $R=1$, then a standard model--like Higgs
boson could be discovered or excluded.
The exclusion potential 
of the $W\phi$ channel at the Tevatron
is summarized in Fig.~\ref{fig:95cl}, while the analogous
discovery potential is described in Fig.~\ref{fig:5sig}.
Fig.~\ref{fig:95cl} shows the 95\% C.L. exclusion limit as a function
of $M_{\phi}$ for LEP2 running at $\sqrt{s}=192$ GeV and 
collecting 150 pb$^{-1}$ of data (dash--dot) and for the Tevatron
 with 30 fb$^{-1}$ (solid),
10 fb$^{-1}$ (long--dash) and 2 fb$^{-1}$ (short--dash).\footnote{This
gives a conservative estimate of the LEP2 reach, which is expected
to be better:  for $\sqrt{s}=200$ GeV and 200 pb$^{-1}$, a maximal
discovery reach of 105 GeV is expected for a standard model--like
Higgs boson.}
The sensitivity of the numerical results for the Tevatron are
demonstrated by the markers, which show the 95\% C.L. exclusion
limit for 2 fb$^{-1}$ 
from a different study that uses the Run I CDF mass resolution
\cite{kuhlmann} (squares) and from the same study with the
backgrounds doubled (triangles).
Fig.~\ref{fig:5sig} shows 5$\sigma$ discovery curves  
as a function of $M_{\phi}$ for LEP2 and the Tevatron 
under the same assumptions as above.
Assuming improved mass resolution from Run I, the Tevatron Run II,
with 2 fb$^{-1}$ of data,
may exclude a 110 GeV standard model Higgs boson at the 95\% C.L.
However, assuming the present mass resolution, the exclusion 
limits for 2 fb$^{-1}$ can
range from about 90 to 102 GeV, assuming the background and
experimental efficiencies are understood within a factor of 2
in the present studies.
With 30 fb$^{-1}$ and improved mass resolution, a SM Higgs boson
of mass 130 
GeV can be discovered within this channel.
 In order to analyze the MSSM case, we
shall consider the case of
 improved mass resolution, since, as we shall show, this will be 
necessary to have good coverage of the MSSM Higgs sector.

The $R$--contour as a function of Higgs boson mass is the starting
point also for a MSSM Higgs boson analysis.
After specifying the parameters of the top squarks,
the function $R$ can be calculated as we scan through
$m_A$ and $\tan\beta$.  
For the MSSM, the numerical results are illustrated 
in Figs.~\ref{fig:max_m}--\ref{fig:amu15}.
Figures \ref{fig:max_m}--\ref{fig:min_p} 
correspond to three common choices of the MSSM parameters.
This set is not exhaustive, but is meant to illustrate the effect of the
stop mixing parameters 
when all sparticles are relatively heavy.  We have taken
the squarks to 
have masses $M_S=1$ TeV.  The higgsino mass parameter is taken to have
the values $\pm 1$ TeV.  Finally, the stop trilinear coupling $A_t$ is chosen
so that the stop mixing parameter is either very small (minimal mixing),
$\tilde{a} \simeq 0$, or,
in the limit of large $m_A$, it maximize the lightest CP--even
Higgs mass (maximal mixing), $|\tilde{a}| \simeq \sqrt{6}$.

In the case of maximal mixing, the radiative corrections to the
${\cal M}_{12}^2$ depend on the sign of 
$\bar\mu \times 
\tilde{a}$. This dependence is obvious in Figs.~\ref{fig:max_m} and 
\ref{fig:max_p}, where
the discovery reach for the cases $\mu = \pm 1$ TeV
and $\tilde{a} = \sqrt{6}$ is displayed.
An increase or
decrease of the effective Yukawa coupling of a standard model--like 
Higgs boson to bottom quarks, for negative or positive values of
$\bar\mu \times  \tilde{a}$, leads
to large variations in the branching ratio to $b$ quarks.  This
has an important impact on
the luminosity required to observe a Higgs boson in this channel.
For Higgs boson masses
in the range 125--130 GeV, as expected for the SM--like Higgs boson
in the limit of large $\tan \beta$ and maximal mixing,
a decrease in the branching ratio to bottom quarks is compensated by
an increase in the branching ratio to $W^{(*)}W^{(*)}$.

From Figs.~\ref{fig:max_m} and 
\ref{fig:max_p} it follows that
 the  dependence on the sign of $\bar\mu \times \tilde{a}$ 
is particularly strong in the
regime of low values of $m_A$, but, as we discussed before, might
become relevant even for relatively large values of $m_A \simeq
{\cal O}(300$ GeV).
The  suppression of the coupling which is obtained for low
values of $m_A$ and large values of $\tan\beta$ leads to a problem
for detecting the heavy Higgs boson in this regime for positive 
values of $\bar{\mu} \times \tilde{a}$ (see Eq. (\ref{suppression})). 

The region of $m_A \simeq 120$ GeV ($m_A \simeq m_{\phi^{SM}}$)
and large $\tan\beta$ is difficult to observe, since this is
the region of maximal mixing 
and $\sin^2(\beta - \alpha) \simeq
\cos^2(\beta - \alpha) \simeq {\cal{O}}(0.5)$. 
Although this limitation can
be overcome with larger luminosity, a window of non--observability 
would remain
for both signs of $\bar\mu \times \tilde{a}$.
Fortunately, if $\tan\beta$ is sufficiently large,
the two CP--even Higgs bosons tend to have similar masses.  Whenever
the mass difference is less than 10 GeV, we have increased the mass window
to include both signals.  
Usually, the number of events within $\pm 2\sigma_M$ of a
given Higgs boson mass is used to quantify the significance 
of any deviation from the expected mean number of events.  
To combine the signals from two Higgs bosons with slightly different
masses, the mass window
 extends from $m_{\phi_{lo}}-2\sigma_M$ 
to $m_{\phi_{hi}}+2\sigma_M$, where $\phi_{lo}$ and $\phi_{hi}$ are the lighter
and heavier of the two Higgs bosons.
Using this procedure, we
can extend the coverage for very large 
$\tan\beta$ and $m_A\simeq m_{\phi^{SM}}$.  However, a window of
non--observability remains for $\tan\beta \simeq 5$, since the
mass difference is large and $\sin(\beta-\alpha)$ is suppressed
in this region.\footnote{If we had not combined the signals, the
region of non--observability for $m_A\simeq m_{\phi^{SM}}$
would have extended to $\tan\beta
\simeq 50~(30)$ in Fig.~3 (4).}
Finally, the region of $m_A \simeq 150$ GeV is easier to observe,
since $\sin^2(\beta - \alpha)$ is already of order one
in this region and the bottom coupling of the lightest CP--even
Higgs boson is strongly enhanced with respect to the standard model
case, implying an increase in the branching ratio of this Higgs boson
into bottom quarks.

In Fig.~\ref{fig:min_p}, the case of minimal mixing 
($\tilde{a}=0$, $\bar\mu=1$)
is displayed. 
In the large $m_A$ limit, the lightest CP--even Higgs mass is of
order 110--115 GeV for moderate or large values of $\tan\beta$ and
hence detectable for luminosities of order 10 fb$^{-1}$. The 
characteristics of this case are similar to the case of maximal
mixing, although, due to the lower values of the lightest CP--even
Higgs mass, lower luminosities are required to cover the large
$m_A$ region and, in addition,  
the window of non--observability for 32 fb$^{-1}$
shrinks to a very small region of parameter space for 
$m_A \simeq m_{\phi^{SM}}$.
For minimal mixing,
the results 
are insensitive to the sign of $\mu$.
This occurs since
the dependence of the Higgs boson mass matrix (and hence
of the CP--even Higgs masses and their couplings to bottom quarks) 
on the sign of $\mu$ arises through the radiative corrections
to the off--diagonal elements, which
are proportional to $\bar{\mu} \times \tilde{a}$ and vanish 
when $\tilde{a} \simeq 0$.  

To illustrate the sensitivity of our results to experimental resolution,
we have constructed Fig.~\ref{fig:min_p_kuhl} 
from the results of another study~\cite{kuhlmann}
which used the present CDF mass resolution.  For worse mass resolution,
the discovery reach at large $m_A$ and near $m_A\simeq m_{\phi^{SM}}$ is
compromised, but the general features are the same.

In Figs.~\ref{fig:amu1} and \ref{fig:amu15},
we present cases in which the stop mixing parameters are such
that the bottom Yukawa coupling of the standard model--like Higgs boson
can be efficiently suppressed in a large region of parameters.  For this,
we have taken values of the mixing parameters $A_t = -\mu = $ 1 and 
1.5 TeV. In these cases, in the limit of large values of $m_A$ and 
for moderate or large values of $\tan\beta$, the 
lightest CP--even Higgs boson mass takes values in the range 115--120 GeV
and 120--125 GeV, respectively. In these two cases, windows of
non--observability appear associated with the suppression of the
bottom Yukawa coupling of the standard model--like Higgs boson, 
i.e. vanishing ${\cal M}_{12}$. 
For $\mu = 1$ TeV, the mass of the standard model--like
Higgs boson  is smaller and the bottom Yukawa coupling cancellation is
more difficult than in the case of $\mu = 1.5$ TeV 
(see Eq.~(\ref{suppression})). Hence, although in the former case
the window of non--observability is small and restricted to small 
values of $m_A$, in the latter case, for large values of $\tan\beta$
the window of non--observability extends up to relatively large values
of the CP--odd Higgs mass.  It would be very interesting to check
if these windows may be efficiently covered by using the $WW$ decay
mode of the Higgs boson~\cite{Wells,Han}.

It is clear from  Figs.~\ref{fig:max_m}--\ref{fig:amu15} that a single detector
at the Tevatron requires
about 30 fb$^{-1}$ for a reasonable 
coverage of the MSSM parameter space, far beyond 
the region already covered by LEP2. 
Other decay channels besides $h,H \rightarrow b \bar{b}$ 
need to be explored to cover specific regions of parameter 
space.
For values
of $M_S \simeq 1$ TeV
and moderate or large values of $\tan\beta$, the standard model--like Higgs
boson tends to be heavier than 100 GeV.  In this case, the LEP reach in the
$Zh$ and $ZH$ channels is highly reduced, and most of the
coverage usually shown is induced by the $hA$ production (since
$HA$ is kinematically limited).\footnote{For $\sqrt{s}=200$ GeV and 
200 pb$^{-1}$, LEP2 can discover a Higgs boson in the $hA$
channel for $m_A<90$ GeV and large $\tan\beta$.}
An essential advantage of the Tevatron is the fact that it can
overcome this kinematic limitation and give
a significant coverage of the $m_A$--$\tan\beta$ plane via the
$Wh$ and $WH$ channels even for large values of $\tan\beta$. 
Obviously, the addition of the $Zh$ and $ZH$
channels would be useful to 
confirm a signal, or to  enhance the possibility of
a discovery with lower integrated luminosity.

\section{Yukawa coupling effects in the large $\tan\beta$ regime.}

In the SM, the coupling of the Higgs boson to $b$ quarks is
proportional to the bottom Yukawa coupling $h_b^{SM} \equiv m_b/v$.
Within the MSSM, 
the effective bottom Yukawa coupling can be quite 
different than in the standard model case. This is due not only to
the dependence on the Higgs mixing angles, discussed above, but also to the
presence of large radiative corrections in the coupling of bottom quarks
and $\tau$ leptons to the neutral components of the Higgs 
doublets, that lead to modifications of 
the relation, Eq. (\ref{tree}), between the bottom 
Yukawa coupling and the running bottom mass~\cite{deltamb,deltamb1,deltamb2}.
To better understand this, it is 
necessary to concentrate on  the properties of the 
large $\tan\beta$ regime. 
In this regime, to a first approximation, only $H_2$ acquires a
vacuum expectation value (here $H_2$ denotes the whole $SU(2)_L$ doublet). 
This means that this Higgs doublet contains
the three Goldstone bosons and a neutral Higgs boson with standard model
like couplings to the electroweak gauge bosons. The other Higgs
doublet $H_1$ does not communicate with the electroweak symmetry
breaking sector and contains, in a first
approximation,  a CP--even and a CP--odd Higgs field, which are almost
degenerate in mass, and a charged Higgs field, 
whose mass differs from $m_A$ only
in a D--term, $m_{H^{\pm}}^2 = m_A^2 + M_W^2$. 
The SM--like Higgs boson acquires a mass 
given by
\begin{equation}
m^2_{\phi^{SM}} \simeq {\cal M}^2_{22} 
\simeq M_Z^2  + {\rm radiative~corrections},
\end{equation}
and its dependence on $m_A$ is suppressed by a $1/\tan^2\beta$ factor
(see Eqs. (\ref{higgsm2l}) and (\ref{matel})). 

Supersymmetric one--loop corrections 
to the tree--level, running bottom quark mass can be significant for
large values of $\tan \beta$ and 
translate directly into a redefinition of the 
relation between the 
bottom Yukawa coupling entering in the production and decay processes and
the physical (pole) bottom mass.
Some of the phenomenological implications of these corrections
have been considered for MSSM Higgs boson decays\cite{sola}.
The main reason why these one--loop corrections
are particularly important 
is that they do not decouple in the limit of a heavy supersymmetric
spectrum.
As mentioned above,
in the supersymmetric limit, bottom quarks only couple to the 
neutral Higgs $H_1^0$. However, supersymmetry is broken and
the bottom quark will receive a small coupling
to the Higgs $H_2^0$ from radiative corrections,
\begin{equation}
{\cal L} \simeq h_b H_1^0 b \bar{b} + \Delta h_b H_2^0 b \bar{b}.
\label{couplings}
\end{equation} 
The coupling $\Delta h_b$ is suppressed by a small loop factor 
compared to $h_b$ and hence, one would be inclined to neglect it.\footnote{
In the above we are explicitly neglecting corrections to the 
$h_b$ coupling of ${\cal{O}} (\Delta h_b$).}
However, once the Higgs doublet acquires a vacuum
expectation value, the running bottom mass receives contributions 
proportional to $\Delta h_b v_2$.  Although $\Delta h_b$ is
one--loop suppressed with respect to $h_b$, 
for sufficiently large values of $\tan\beta$ ($v_2 \gg v_1$)
the contribution to the bottom quark mass of both terms in 
Eq. (\ref{couplings}) may be comparable in size. This induces a
large modification in the tree level relation, Eq.~(\ref{tree}),
\begin{eqnarray}
m_b = h_b v_1 (1+\Delta(m_b)),
\label{yukbmass}
\end{eqnarray}
where $\Delta(m_b) = \Delta h_b \tan\beta/h_b$.

The function $\Delta(m_b)$ contains two main
contributions, one from a bottom squark--gluino loop 
(depending on the two bottom squark masses $M_{\tilde b_1}$
and $M_{\tilde b_2}$ and the gluino mass $M_{\tilde g}$) and another one
from a
top squark--higgsino loop (depending on the two top squark masses
$M_{\tilde t_1}$ and $M_{\tilde t_2}$ and the higgsino mass parameter
$\mu$).  The explicit form of $\Delta(m_b)$ at one--loop can be approximated
by computing the supersymmetric loop diagrams at zero external momentum
($M_S \gg m_b$) and is given 
by~\cite{deltamb,deltamb1,deltamb2}:
\begin{center}
\begin{eqnarray}
\Delta(m_b) \simeq {2\alpha_3 \over 3\pi} M_{\tilde g}\mu\tan\beta~I(M_{\tilde b_1},
M_{\tilde b_2},M_{\tilde g})
 + {Y_t \over 4\pi} A_t\mu\tan\beta~I(M_{\tilde t_1},M_{\tilde t_2},\mu),
\label{deltamb}
\end{eqnarray}
\end{center}
where $\alpha_3=g_3^2/4\pi$, 
$Y_t= \frac{h_t^2}{4\pi}$, 
and the function $I$ is given by,
\begin{center}
\begin{eqnarray}
I(a,b,c) = {a^2b^2\ln(a^2/b^2)+b^2c^2\ln(b^2/c^2)+c^2a^2\ln(c^2/a^2) \over
(a^2-b^2)(b^2-c^2)(a^2-c^2)},
\end{eqnarray}
\end{center}
and is positive by definition.
Smaller contributions to $\Delta(m_b)$ have been neglected for the
purpose of this discussion.
It is important to remark that these effects are just a manifestation
of the lack of supersymmetry in the low energy theory and, hence,
$\Delta(m_b)$ does not decouple 
in the limit of large values of the supersymmetry breaking masses. Indeed,
if all supersymmetry breaking parameters (and $\mu$)
are scaled by a common factor, the correction
$\Delta(m_b)$ remains constant.

Similarly to the bottom case, the relation between $m_\tau$ and 
the $\tau$ lepton Yukawa coupling $h_\tau$ is modified:
\begin{eqnarray}
m_\tau = h_\tau v_1 (1+\Delta(m_\tau)).
\end{eqnarray}
The function $\Delta(m_\tau)$ contains a contribution from a tau slepton--
bino loop (depending on the two stau masses $M_{\tilde \tau_1}$
and $M_{\tilde \tau_2}$ and the bino mass parameter $M_{1}$) and a
tau sneutrino--chargino loop (depending on the tau sneutrino mass
$M_{\tilde \nu_\tau}$, the wino mass parameter $M_{2}$ and $\mu$).
It is given by the expression~\cite{deltamb1,deltamb2}:
\begin{eqnarray}
\Delta(m_\tau) = {\alpha_1 \over 4\pi} M_1\mu\tan\beta I(M_{\tilde\tau_1},
M_{\tilde\tau_2},M_1) + {\alpha_2 \over 4\pi} M_2\mu\tan\beta \;
I(M_{\tilde\nu_\tau},M_2,\mu),
\end{eqnarray}
where $\alpha_1=\frac{g_1^2}{ 4\pi}$, $g_1$ is the $U(1)$ hypercharge coupling,
$\alpha_2=\frac{g_2^2}{4\pi}$, $g_2$ is the $SU(2)$ weak isosopin coupling.

Since corrections to $h_\tau$ are proportional to $\alpha_1$ and
$\alpha_2$, they are  expected to be smaller than 
the corrections to $h_b$.
Although the precise values of $\Delta(m_b)$ 
and $\Delta(m_{\tau})$ are model dependent,  the 
leading term in the tau mass corrections  has a factor 
$M_2\mu I(M_{\tilde\nu_\tau},M_2,\mu) \propto 
M_2\mu /(\max(m_{\tilde{\nu}_{\tau}}^2,M_2^2,\mu^2) \leq 1$, and hence,
for $\tan\beta \simlt 50$, 
$\Delta(m_{\tau}) < 0.15$.
In the following, we consider the impact of the bottom mass 
corrections assuming $\Delta(m_{\tau}) \ll \Delta(m_b)$, and
using the expression $\Delta(m_b) = K\tan\beta$ to
parametrize possible radiative corrections.
Since the
value of $\alpha_3$ at the scale $M_S$ is
of order 0.1, and if all soft supersymmetry breaking parameters and
$\mu$ are of order of 1 TeV, the coefficient $K$ can have either
sign and will be of order $|K| \simeq 0.01$. One can also consider
cases in which the bottom mass corrections are highly suppressed.
This happens naturally in the case of approximate
$R$ and Peccei--Quinn symmetries in the theory, which make the value
of the gaugino masses and the stop mixing parameters much lower
than $M_S$~\cite{deltamb}. 

It is instructive to return to
the couplings of the lightest and heaviest CP--even Higgs bosons and of
the CP--odd Higgs boson to bottom
quarks. The CP--odd Higgs boson coupling to bottom quarks is given by
\begin{equation}
{\cal L} = - i {h_b}^{CP} A \bar{b} \gamma_5 b
\end{equation}
with
\begin{equation}
h_b^{CP} = 
h_b \sin\beta + \Delta h_b \cos\beta \simeq h_b \sin\beta =
{m_b\over (1 + \Delta(m_b)) v}\tan\beta.
\label{bhCP}
\end{equation}
Using Eqs. (\ref{couplings}) and (\ref{mixings}), together
with the relation of the bottom Yukawa coupling to the bottom mass,
Eq. (\ref{yukbmass}), it is easy to show that the effective 
couplings of the CP--even Higgs bosons, $\bar{h}_b$ and $\tilde{h}_b$,
\begin{equation}
{\cal L} = \bar{h}_b b \bar{b} h + \tilde{h}_b b \bar{b} H 
\end{equation}
are approximately given by
\begin{equation}
\bar{h}_b  \simeq -\frac{m_b \sin\alpha}{v \cos\beta}\
\left[ 1 - \frac{\Delta(m_b)}{1 + \Delta(m_b)} \left( 1 + \frac{1}
{\tan\alpha \tan\beta} \right) \right],
\label{barhb}
\end{equation}
\begin{equation}
\tilde{h}_b \simeq \frac{m_b \cos\alpha}{v \cos\beta}
\left[ 1 - \frac{\Delta(m_b)}{1 + \Delta(m_b)} 
\left( 1 - \frac{\tan\alpha}{\tan\beta} \right) \right].
\label{tildehb}
\end{equation}
~\\
The value of $\Delta(m_b)$ 
in eq. (\ref{deltamb}) is defined
at the scale $M_{S}$,
where the sparticles are decoupled. The $ h_b$ and $\Delta h_b$  couplings  
should be computed at that scale, and run down with their respective 
renormalization group equations to the scale $m_A$, where the 
relations between the couplings of the bottom quark  to the neutral Higgs 
bosons and the running bottom quark mass, Eqs. (\ref{bhCP}), (\ref{barhb}), 
and (\ref{tildehb}) are defined.
In the present study we have 
defined the running bottom mass at the scale $m_A$ as a function of 
$m_b(m_b) \simeq$ 4.25 GeV while using two--loop renormalization group 
equations in the effective standard model theory at scales $Q < m_A$. 
The above procedure leads to 
 a consistent definition of the bottom quark couplings to
the Higgs bosons when the three neutral Higgs boson
masses are of the same order. 
For large values of the CP--odd Higgs boson mass, instead, $\bar{h}_b$ must be 
evolved with SM renormalization group equations from $m_A$ down to $m_h$.
The definition of the couplings of the bottom quark to the Higgs bosons
at the scale of the 
corresponding Higgs boson mass 
is  chosen to take into account the bulk of the QCD correction.
Indeed, it
is known that this choice of scale represents well
the bulk of the QCD corrections to the Higgs boson decay into quarks
and gluons~\cite{CARZER}. However, 
for the production process $b \bar{b} \phi$,
a complete study of the NLO effects remains necessary to get
a definitive estimate of the Tevatron reach in this production channel.

It is interesting to study different limits of the above couplings.
For large $m_A \gg m_{\phi^{SM}}$, 
the lightest CP--even Higgs boson should behave like
the SM particle. This is fulfilled since, in this limit
$\cos\beta \simeq -\sin\alpha$ and $\sin\beta \simeq \cos\alpha$. Hence,
$\bar{h}_b = m_b/v$, which is the standard coupling. In the
same limit, $\tilde{h}_b \simeq h_b \sin \beta (1 + {\cal O}
(\Delta(m_b)/\tan^2\beta))$.
Even in the presence of 
radiative corrections to the bottom quark couplings,
the heaviest CP--even Higgs boson coupling is approximately equal
to the CP--odd one. When $m_A$ starts approaching $m_{\phi^{SM}}$, the 
above relations between the angles $\alpha$ and $\beta$ are slightly
violated. 
Due to the large $1/\cos\beta$ factor appearing in the definition
of the Yukawa coupling, Eq. (\ref{barhb}),
a small departure from the above relations can induce large
departures of the coupling $\bar{h}_b$ with respect to the standard model
value.
For $m_A \ll m_{\phi^{SM}}$, instead, $\sin\beta \simeq -\sin\alpha \simeq 1$.
The lightest Higgs boson coupling is $\bar{h}_b \simeq 
 h_b \sin \beta (1 + {\cal O}(\Delta m_b/\tan^2\beta))$, while, as 
happens for vanishing values of the bottom mass corrections, 
the coupling
of the heaviest CP--even Higgs boson may become highly non--standard.

As discussed in Sec. 2, in the large $\tan\beta$
regime, the off--diagonal elements of the mass
matrix can receive large radiative corrections
with respect to the tree--level value, $\left({{\cal M}^2_{12}}\right)^{tree} 
\simeq -(m_A^2 + M_Z^2)/\tan\beta$. When both $\Delta(m_b)$ and
$\Delta(m_{\tau})$ are small, the coupling of the standard model--like 
Higgs boson to bottom quarks and $\tau$ leptons vanishes for
vanishing $\sin 2 \alpha$ (see 
Eq. (\ref{sin2alpha})). The reason for this cancellation 
when $\sin2\alpha = 0$ is that the standard model--like Higgs boson
becomes a pure $H_2^0$ state, which does not couple to bottom 
quarks and $\tau$--leptons at tree level. 
If the bottom and $\tau$ mass corrections are
large, however, the bottom and $\tau$ couplings do not cancel for
$\sin2\alpha = 0$, but are just given by $\Delta h_b$ and
$\Delta h_{\tau}$, respectively. Indeed, from Eq. (\ref{barhb}) 
(Eq. (\ref{tildehb})), in the
limit $\sin\alpha = 0$ ($\cos\alpha = 0$), the bottom coupling
is given by 
\begin{equation}
\bar{h}_b (\tilde{h}_b) = \frac{m_b}{\sin\beta v} \times \frac{\Delta(m_b)}
{(1 + \Delta(m_b))}  \equiv \Delta h_b.
\label{barhb2}
\end{equation}
In this limit,
the coupling to bottom quarks is much smaller than the standard model 
coupling only if $|\Delta(m_b)| \ll 1$. 
A similar expression to Eq.~(\ref{barhb2}) holds for the $\tau$ lepton coupling.

For values of 
$\Delta(m_b)$ of order 1, however, a strong suppression of the bottom 
coupling $\bar{h}_b$ can still occur for slightly different values of the Higgs
mixing angle $\alpha$, namely
\begin{equation}
\tan\alpha \simeq \frac{\Delta (m_b)}{\tan\beta}.
\end{equation}
Under these conditions,
\begin{equation}
\bar{h}_{\tau} = \frac{m_{\tau}}{v \sin\beta} \left( 
\frac{ \Delta(m_{\tau}) - \Delta(m_b)}{ 1 + \Delta(m_{\tau})} \right),
~~~~~\bar h_b=0.
\end{equation}
A similar expression is obtained for the coupling $\tilde{h}_{\tau}$
in the case $\tilde{h}_b = 0$.
Hence, if $\tan\beta$ is very large and $\Delta(m_b)$ is of order
one, the $\tau$ Yukawa coupling may {\it not} be strongly suppressed with
respect to the standard model case and can provide the {\it dominant}
decay mode for a standard model--like Higgs boson.

To recapitulate, the cancellation in the off--diagonal elements of the mass
matrix can lead to a strong suppression of the
standard model--like Higgs boson coupling to bottom quarks and $\tau$
leptons. In general, this implies a sharp increase of the 
branching ratio of the decay of this Higgs into gauge bosons
and charm quarks. However, for very large values
of $\tan\beta$ and values of the bottom mass corrections
$\Delta (m_b)$ of order one, the branching ratio of the decay into
$\tau$ leptons may increase in the regions in which the bottom
quark decays are strongly suppressed.

\section{Higgs phenomenology with large $\tan\beta$ corrections.}

\subsection{$W\phi$ process}

The finite corrections to the bottom Yukawa coupling are
important in defining the exact regions for which the
bottom Yukawa coupling is suppressed.
Depending on the sign of the bottom mass
corrections and on the specific region of supersymmetric parameter space, 
important increases or decreases in coverage may occur with respect 
to the case of  $\Delta(m_b)=0$. 
For large $m_A$, the coupling of the lightest
CP--even Higgs boson is only slightly affected by the presence
of $\Delta(m_b)$, and
these corrections will not affect the
discovery potential.
The only exception is when the negative
contributions to the ${\cal M}_{22}^2$ matrix elements, proportional
to $h_b^4$, become relevant (see Eq. (\ref{matel})). 
For low values of $m_A$, instead, the bottom mass corrections
might have an important impact in the discovery and exclusion
reach for a given choice of parameters.

Figures~\ref{fig:max_p_m005}--\ref{fig:max_p_p01} 
show the impact of the bottom mass corrections on
the discovery reach of the CP--even Higgs bosons in the
$W\phi$ channel for the case of maximal mixing, 
$\mu>0$, and both signs
of the bottom mass corrections, assumed to be given by
$\Delta m_b = K \times \tan\beta$, with $K = \pm 0.05$ and
$\pm 0.01$.  Since the 
Higgs sector parameters depend only on the size of the mixing
parameters and on the sign of $\bar{\mu}
\times \bar{A}_t$, while the bottom mass corrections depend also
on the sign of $\bar{\mu} \times M_{\tilde{g}}$, one can have
either sign for the bottom mass corrections, for fixed
values of the stop mixing parameters.
Although the most generic features of the discovery reach plots
are not changed by the presence of the bottom mass corrections,
positive bottom mass corrections tend to reduce the
bottom Yukawa couplings and increase the luminosity needed
for a Higgs boson discovery. The opposite happens in the presence of
negative mass corrections. Observe that, for values of $K = -0.01$,
there is an improvement of the 
discovery reach at very large values of $\tan\beta$.
This improvement is related to a decrease in the lightest CP--even 
Higgs boson
mass induced by the negative corrections to the ${\cal M}_{22}^2$ 
Higgs mass squared matrix elements, proportional to $h_b^4$, which
become enhanced for large values of $\tan\beta$ and negative values
of $K$. As we shall discuss below, in minimal supersymmetry breaking
models, the bottom mass corrections tend to be positive,
and hence the reach of the Tevatron is negatively affected.
The same plots for the case of minimal mixing do
not show as much sensitivity.

The reason why positive bottom mass corrections suppress the
reach of the Tevatron collider can be easily understood by
studying the behavior of the effective bottom Yukawa couplings,
Eqs. (\ref{barhb}) and (\ref{tildehb}). Indeed, for $m_A \ll m_{\phi^{SM}}$,
since $\tan\alpha/\tan\beta<0$, the expression between parenthesis
in  Eq. (\ref{tildehb}) 
is positive.  For a fixed value of the
angles $\alpha$ and $\beta$, a positive $\Delta(m_b)$
tends to reduce the value of $\tilde{h}_b$. For $m_A \gg m_{\phi^{SM}}$,
instead, since $\tan\alpha \times \tan\beta <0$, 
the effect of the bottom mass
corrections on the value of $\bar{h}_b$, Eq. (\ref{barhb}), 
depends on whether  $|\tan\alpha \times \tan\beta|$
is greater or less than 1. In the cases displayed in Figs.~9--12, this
factor is always larger than one and a positive bottom mass
correction reduces the value of $\bar{h}_b$.

\subsection{$b\bar b\phi$}

The Yukawa coupling corrections discussed above affect 
the associated production of a Higgs boson
with $b$ quarks, where the Higgs boson subsequently decays to 
a heavy flavor final state.  
Indeed, the cross section times branching ratio for this process
at an $h_1h_2$ hadron collider satisfies
\begin{eqnarray}
\sigma(h_1h_2\to b\bar b \phi(\to b\bar b)+X) \propto 
\hat{h}_b^2  
\times \hat{B}_b ,
\label{bbbb}
\end{eqnarray}
and
\begin{eqnarray}
\sigma(h_1h_2\to b\bar b \phi(\to\tau^+\tau^-)+X) \propto 
\hat{h}_b^2  
\times \hat{B}_\tau ,
\label{corr_bbtt}
\end{eqnarray}
where $\hat{h}_b=h_b^{CP},\bar{h_b},$ or $\tilde{h_b}$ depending
on $\phi$, and
$\hat{B}_b$ and $\hat{B}_\tau$ are the corresponding branching ratios
of the $\phi$ decay into bottom quarks and $\tau$ leptons, which
are computed using the modified
couplings $\hat{h}_b$ and $\hat{h}_\tau$. In general,
while the four $b$ final state, Eq. (\ref{bbbb}),
is strongly affected by $\Delta(m_b)$, the $b\bar b \tau^+\tau^-$ final
state~\cite{drees}, Eq. (\ref{corr_bbtt}), is only mildly affected 
due to a cancellation of the dependence of the production cross 
section times branching ratios on this factor.

\subsubsection{On the CP--even Higgs boson masses  at large values of
{\large $\tan\beta$}}

One interesting feature of the large $\tan\beta$ regime is that
the CP--odd and one of the
two CP--even Higgs bosons have similar masses and couplings.  One might
be tempted to take the signal from $b\bar b A$ production and
decay and double it to account for the other non--SM--like Higgs
boson.  However, this approximation is optimistic, and not
necessary.
For example, when both 
$|\bar{\mu}|$ and $| \tilde{a} | \simgt {\cal{O}}(1)$,this might be 
a poor approximation~\cite{Hempfling}. Indeed, the
CP--even Higgs boson with similar properties to the CP--odd one
has a mass approximately equal to 
${\cal M}_{11}$, with the form
\begin{equation}
{\cal M}_{11}^2 \simeq m_A^2 - \frac{m_t^4}{16 \pi^2 v^2}\bar{\mu}^2
\tilde{a}^2,
\end{equation}
where we have omitted the two-loop corrections, Eq. (\ref{matel}).
A particularly interesting case to analyze is the maximal mixing case, 
$\tilde{a}^2 = 6$, when
the radiative corrections to
$m_{\phi^{SM}}^2$ are maximized and 
${\cal M}_{12}^2$ receives only small radiative
corrections for moderate or large values of $m_A$. 
For small mass differences compared to the
average mass, one gets approximately
\begin{equation}
m_A - m_h \simeq \frac{3 m_t^4}{8 \pi^2 v^2} 
\frac{\bar{\mu}^2}{2 m_A} \;\;\;\;\;\;\;\;\;\;\;\;\;\;\;\;\;\;
(\tilde{a}^2 = 6).
\end{equation}
For $|\bar{\mu}| \simgt 1$, 
both CP--even Higgs boson masses can be significantly different from
the CP--odd one.
Fig.~\ref{fig:dmass} shows the minimal and maximal
mass difference  of the CP--even Higgs boson mass $m_h$
($m_H$) with the CP--odd Higgs boson mass, for values of $\sin^2(\beta-\alpha) < 0.5$
($\sin^2(\beta-\alpha) > 0.5$), in the maximal mixing case and
$|\bar{\mu}| = 1$.  A scan
was performed over values of $m_A > 80$ GeV. 
As is clear from the above expression, the maximal and minimal mass
differences are obtained for the minimal and maximal values of $m_A$ chosen.
For instance, for values of $m_A \simeq 80$ GeV, $\tilde{a} = \sqrt{6}$
and $\bar{\mu} = \pm 1$, one obtains a mass difference
of about 5 GeV for large $\tan\beta$, 
which coincides with the results presented in the figure.
Had we scanned over lower values of $m_A$, the mass difference would have
increased. In our analysis we consider the separate signals from $A$ and the
CP--even like Higgs boson with similar masses and couplings as the CP-odd Higgs
boson.

\subsubsection{Simulation of $b\bar b(\phi\to b\bar b)$ signal and backgrounds}

The numerical results for the $b\bar b\phi$ process are based on the 
study of a generic neutral Higgs $\phi$ (with production and decay properties
of the CP--odd Higgs boson).\footnote{This process was first considered at the
Tevatron based on a 3 jet analysis \cite{gunion}.  Later, it was reconsidered
based on a 4 jet analysis \cite{tait}.  The modified results
of Ref.~\cite{tait} presented in Ref.~\cite{nafta}
are now in general
agreement with our results when comparable.}
First, we consider the four $b$--quark final state from
the decay $\phi(\to b\bar b)$.  
We performed a parton--level simulation based 
on {\tt Madgraph} \cite{madgraph} matrix elements for the processes
$p\bar p\to b\bar b \phi(\to b\bar b)+X$, 
$\to b\bar b Z(\to b\bar b)+X$, and $\to b\bar b b\bar b$ (QCD).
All matrix elements are evaluated at leading--order, using
leading--order $\alpha_s$, leading--order parton distribution
functions (CTEQ3L), and a common scale $\sqrt{\hat{s}}/2$, where
$\sqrt{\hat{s}}$ is the partonic center--of--mass 
energy.\footnote{The scale dependence of these results
is estimated to be 25\%\cite{nafta}.  
In the same study, the reach of a 3 or 4 $b$--tag analysis
is estimated to be the same.}
The Higgs
boson and $Z$ boson resonances are treated in the narrow width
approximation.
Next--to--leading order QCD corrections to these processes are
expected to be important, but they have not yet been calculated.
Since the  $b\bar b$ production cross
section at the Tevatron at NLO is almost doubled from the LO result,
as a  first estimate the signal and backgrounds in this study are multiplied
by a factor of 2\cite{kfactor}.  This assumption will need to be considered
in more detail elsewhere.
When Gaussian statistics apply, this increases
the significance ($S\over\sqrt{B}$) of a signal by $\sqrt{2}$.
We assume that four $b$--quark tags are necessary to reduce
backgrounds (so we do not consider $Z(\to b\bar b)jj$ or $b\bar bjj$ backgrounds), 
and that the four--tag efficiency can be described by an
overall factor  of $(.45)^2\simeq .2$.\footnote{This estimate is
based on the double tagging efficiency of Ref.~\cite{kuhlmann}.
The actual efficiency will require a detailed analysis.}
We also assume that the efficiency for
triggering on a four $b$ final state is unity.  

The signal is defined by the following cuts:
\begin{itemize}
 \item 4 $b$ partons with ${p_T}_i>20$ GeV, i=1,4, and $|\eta^i|<2$,
 \item $\max({p_T}_i)>40$ GeV
 \item $R_{ij}>0.8$, where $R_{ij}=\sqrt{(\phi_i-\phi_j)^2+(\eta_i-\eta_j)^2}$.
\end{itemize}
For events that satisfy these cuts, the distribution of
all invariant mass combinations $m_{ij}=\sqrt{(p_i+p_j)^2}$ is
constructed.  
We use a mass resolution of $\sigma_M/M=.08$ above 100 GeV, and
$\sigma_M=8$ GeV below.  
We use the procedure outlined earlier 
to combine the signals from two Higgs bosons which are close in
mass.

For all Higgs boson masses considered in this study, the QCD production
of four $b$--quarks is the dominant background.  The cross section
for this process is proportional to
$\alpha_s^4$, and, hence, is sensitive to the choice of scale.
Therefore, an absolute prediction of the event rate after cuts has a
large uncertainty.
By studying all $m_{ij}$ combinations, it is possible to define
a smooth background distribution and determine the overall normalization
from the data using sidebands.
If, instead, one chooses the $m_{ij}$ combination closest to a 
hypothesized Higgs boson mass, then a distribution similar to
the signal is sculpted from the
background, and it becomes problematic to assess the significance of
a mass peak.
 
There may be optimal cuts to increase the significance for heavier
Higgs boson masses.  For $m_\phi>100$ GeV, we include the
additional requirement that $|\cos\theta^*|<0.8$, where $\theta^*$
is the polar angle distribution in the rest frame of the $b\bar b$ pair
that best reconstructs to $m_\phi$.

\subsubsection{Results}


The 95\% C.L. exclusion and 5$\sigma$ discovery 
potentials of the $b\bar b b\bar b$
channel are illustrated in Fig.~\ref{fig:bbbb} for different total 
integrated luminosities and for the case of minimal mixing.
These results imply that a CP--odd Higgs boson (and its partner
with similar properties) can be discovered with 30 fb$^{-1}$ of data
at large $\tan\beta$ if $m_A \simlt 200$ GeV.  For the same integrated
luminosity, an exclusion contour can cover from $m_A \simeq 80$ GeV
and $\tan\beta\simeq 15$ up to $m_A\simeq 300$ GeV and large
$\tan\beta \simeq 50$.  However, these conclusions assume vanishing SUSY
corrections to the bottom mass ($\Delta(m_b)=0$).
Fig.~\ref{fig:bbbb1}  shows the sensitivity of
these results (for maximal mixing and $\mu>0$)
to SUSY corrections at large $\tan\beta$.  
The lines in Fig.~\ref{fig:bbbb1} show the variation of the 
$5\sigma$ discovery contour with 30 fb$^{-1}$
for $\Delta m_b= K \times \tan \beta$, with 
$K$ = $\pm .005$ and $\pm .01$. 
There is a similar variation in the exclusion contours.
Clearly, it is difficult to make a definitive statement about the
reach of the Tevatron in the $b\bar b b\bar b$ final state with limited
knowledge of the size of the bottom mass corrections, $\Delta(m_b)$, which 
depend on the sparticle masses and mixings.  
It is important to 
realize that, for negative values of $K$ and large values of
$\tan\beta$, the cross section increases due to a large increase in
the  bottom Yukawa coupling $h_b$, 
but this Yukawa coupling may become too large
to make a perturbative analysis possible. This would happen, for
instance for values of $|K| \simgt 0.015$ and $\tan\beta \simgt 50$. 

One of the generic conclusions of this study is that, in the
large $\tan\beta$ regime, this process may be useful to test regions
of parameter space which will remain otherwise
uncovered. In general, however, the reach in the $m_A$--$\tan\beta$ 
plane is reduced to a relatively small region
for values of $m_A$ of the order of the weak scale. 
The exact discovery and exclusion potentials depend strongly
on 
the finite SUSY corrections to the bottom
mass, which can be very large.

\subsubsection{Simulation of $b\bar b\phi(\to\tau\tau)$ signal and backgrounds}

We have also considered the possibility of detecting the Higgs boson
decays into $\tau^+\tau^-$.\footnote{This channel was
previously considered for Run I~\cite{drees}. However,
due to a numerical error, the reach in $\tan\beta$ was
largely overestimated in that work~\cite{thanksdress}.}
Since $\Delta(m_\tau)$ is expected to be small, 
the $b\bar b\tau^+\tau^-$ channel is not very sensitive to SUSY--induced,
large $\tan\beta$ corrections.
This follows from
Eq.~(\ref{corr_bbtt}) in the limit that the total width of the Higgs
boson is dominated by the $b\bar b$ partial width.
We present a preliminary study here, but more work needs to be done
to understand the feasibility of this channel.
With our present understanding, Fig.~\ref{fig:bbbb} shows that 
for $m_\phi<$ 120 GeV, the reach of this channel is comparable to or even 
slightly better than the $b\bar b b\bar b$ channel, and 
one may expect  much room for improvement.

To study  this process at the Tevatron, we considered
only the $\tau_1(\to \ell+X)$ and $\tau_2(\to j+X)$ decays of the
$\tau_1\tau_2$ pair, where $\ell=e$ or $\mu$.  
This combination yields a triggerable lepton,
a narrow jet, and two $b$--quarks in the final state.  
It also has the largest branching ratio.
The 
physics backgrounds are assumed to be $b\bar b Z(\to\tau\tau)$
and $t\bar t$ production.
The signal $b\bar b\phi(\to \tau\tau)$ and the background
$b\bar bZ(\to \tau\tau)$ were simulated in a similar manner 
as before and increased by a factor of 2 to account for 
higher--order corrections.
In addition, the $\tau$ polarization information was included 
for all $\tau$ decays
through the 
{\tt Tauola} Monte Carlo program \cite{tauola}.  
The $t\bar t$ background was simulated using {\tt Pythia 6.1} \cite{pythia}
with
the default settings, and forcing $W$ decays into $e,\mu$ or $\tau$.
For this analysis, the $b$--parton is treated as a $b$--jet, but the
$\tau$--jet is constructed from final state particles ($\pi$, $K$, etc.).
The
$\slashchar{E}_T$ is constructed from the real neutrinos from
$W$ boson and $\tau$ decays.

Because the number of backgrounds is smaller than the previous case,
and the signal and $Z$ background have similar characteristics,
it is assumed that the acceptance cuts can be looser.
The basic cuts are:
\begin{itemize}
  \item 2 $b$ partons with ${p_T^b}>10$ GeV, $|\eta^b|<2$.
  \item 1 $e$ or $\mu$ with ${p_T^\ell}>10$ GeV, $|\eta^\ell|<2$.
  \item 1 $\tau$--jet with ${p_T^j}>15 $ GeV, $|\eta^j|<2$.
  \item $R_{ij}>0.6$, where $i,j$ sum over $b$'s, $\ell$ and $\tau$--jet.
\end{itemize}
After these cuts, the $t\bar t$ events produce the largest background.
However, the jets and leptons from $t$--quark decays are much harder
and produce much more $\slashchar{E}_T$ than the typical signal event.
The further cuts $p_T^b<60$ GeV and $\slashchar{E}_T<$ 80 GeV are imposed
to reduce this background.
For the final numbers, the CDF $\tau$--jet reconstruction efficiency ranging
from approximately .3 to .6 is used\cite{cdf_leptoquark},
as well as a double $b$--tag efficiency of .45 and a triggering
efficiency of unity.

The signal is defined by a simple counting experiment, without
reference to a mass window.  
Several possible improvements could greatly increase the
potential of the $b\bar b\tau\tau$ signal.  
First, with adequate $\slashchar{E}_T$
resolution, a mass peak can be partially reconstructed.  This could
distinguish the signal from the background for $m_\phi \gg M_Z$.
Second, the second largest branching ratio for the decay
of a $\tau$ pair is when both $\tau$'s decay to jets.  While this channel
would greatly enhance the signal, it requires a detailed background and
triggering analysis beyond the scope of this work.

\section{Constraints from the decay $b\rightarrow s \gamma$}

As shown above, the couplings of the CP--even Higgs bosons to bottom
quarks depend strongly on $\Delta(m_b)$.
In the MSSM framework, positive or negative corrections are possible.
However, in some specific models, 
the sign of the correction is correlated with 
the supersymmetric contribution to the amplitude of
the decay process
B($b\to s\gamma$), which, at large values of $\tan\beta$, is
proportional to $A_t \times \mu\tan\beta$ \cite{Masiero2}. 
This is the case,
for instance, in the minimal supergravity model, with unification
of the three gaugino masses. 
For moderate
and large values of $\tan\beta$~\cite{Costas,deltamb1},
\begin{equation}
A_t \simeq \frac{A_0}{4} - 1.5 M_{1/2},
\end{equation}
where $A_0$ and $M_{1/2}$ are the boundary conditions for $A_t$ and
the gaugino masses, respectively, at the Grand Unified Theory (GUT) scale. 
Unless $A_0 \simgt 6M_{1/2}$, we have
$A_t \times M_{\tilde g} < 0$. 
Moreover,
it has been shown~\cite{deltamb1} that
unless $A_0$ and $m_0$ are much larger than $M_{1/2}$, the
expression for $\Delta(m_b)$ in Eq.~(\ref{deltamb})
is dominated by the gluino contributions, which are proportional
to $\mu \times M_{\tilde g}$. The top squark--induced 
corrections, proportional to the trilinear parameter $A_t$, 
are smaller than the gluino--induced ones and
tend to reduce the total bottom mass corrections.
Hence, the sign of the bottom mass
corrections is determined by the gluino--sbottom loop contribution, which
is  opposite in sign to the  chargino--stop corrections
to the $b \rightarrow s \gamma$ decay rate.
Cancellation of the positive contribution of the charged Higgs boson
$H^+$ to B$(b\to s\gamma)$ 
requires $A_t \times \mu < 0$, so that in these models the
bottom mass corrections $\Delta(m_b)>0$. Positive corrections to
the bottom mass ($\Delta(m_b)>0$) reduce
the effective bottom Yukawa coupling 
with respect to the tree level value, Eq. (\ref{tree}),
which reduces the discovery and exclusion potential 
of a Higgs boson in the 
$b\bar b b\bar b$ final state at
the Tevatron collider (see Fig.~\ref{fig:bbbb1}).  As discussed above, positive
mass corrections have also a negative effect on the 
exclusion and
discovery potential of a CP--even Higgs boson in the $W\phi$ channel
(see Figs.~\ref{fig:max_p_p005} and \ref{fig:max_p_p01}).

In general, in the absence of flavor violating couplings of the 
down squarks to gluinos, 
the $B(b \rightarrow s \gamma)$ constraint 
on the possible values of $m_A$ and the stop mass parameters
becomes strong for large values of $\tan\beta$. 
For low values of the CP--odd Higgs boson mass, positive values of
$\mu \times A_t$ are disfavored by the data.
Even for negative values of
$\mu \times A_t$, when $m_A \simeq M_Z$,
the suppression induced by
the chargino--stop contributions tends to be too small
to cancel the large charged Higgs boson enhancement,
unless $|\bar{A}_t \times \bar{\mu}| \times \tan\beta$ becomes  large.
This cannot be achieved by pushing the value of the $\mu$ parameter
to values larger than $M_S$, since this would 
increase the chargino masses and lower the loop effect.
Hence, large values of $\tan\beta$ and $|A_t|$ are preferred.
 

The above mentioned constraints on the stop mass parameters
for low values of $m_A$ can be avoided in the presence of a
non--trivial down squark flavor mixing.
In particular, non--negligible mixing parameters
between the second and third generation of down squarks can
contribute to the $b \rightarrow s \gamma$ decay rate via gluino--squark loop
induced processes. 
For instance, if the gluino contributions
were the only ones leading to the $b \rightarrow s \gamma$
decay rate, the branching ratio would be given by~\cite{Masiero}
\begin{equation}
B(b \rightarrow s \gamma) \simeq \frac{2 \alpha_3^2 \alpha_{\rm em}}
{81 \pi^2 M_S^4} m_b^3 \tau_B M_{\tilde{g}}^2 F^2(x) 
\left|\frac{\Delta_{23}^d}{M_S^2}\right|^2,
\label{btosg}
\end{equation} 
where $\Delta_{23}$ is the value of the off--diagonal 
sbottom--sstrange left--right term in the down squark squared
mass matrix (which we assume to be equal to the right-left one), 
$\tau_{B} \simeq 1.5 \times 10^{-12} s$, 
$x= M_{\tilde g}/M_S$ and
\begin{equation}
F(x) = 4 \left( \frac{ 1 + 4 x - 5 x^2 + 4 x \ln(x) + 2 x^2 \ln(x)}
{ 8 ( 1- x)^4} \right).
\end{equation}
This expression
ignores the potentially relevant contributions coming
from a left--left down squark mixing~\cite{Masiero,Blazek}.
Observe that, for $M_{\tilde g} = {\cal O}(M_S)$
and $M_S = {\cal O}$(1 TeV), the contribution of the gluino
mediated diagram to the $b \rightarrow s \gamma$ branching ratio 
is of the order of $(\Delta_{23}^d/M_S^2)^2$. Hence, even a
small left--right mixing, $\Delta_{23}^d$, 
of order of $10^{-2}\times M_S^2$ can induce important
corrections to the amplitude of this decay rate. 
It is straightforward
to show that these low  values of $\Delta_{23}$ 
do not have an immediate 
impact on the Higgs boson sector. 

In the presence of non--trivial flavor mixing in the
down squark sector, large corrections to the amplitude of the
$b \rightarrow s \gamma$ decay rate may be induced. These
corrections may be helpful in determining values of $B(b \rightarrow s \gamma)$
consistent with experimental data for small values of $m_A$ and/or
positive values of $\mu \times A_t$.
For the above reasons, in our presentation,
we have decided to keep the results  for both signs of 
$\bar{A}_t \times \bar{\mu}$. The reader must keep in mind, however,
that positive values of this parameter for low values of $m_A$
would imply a more complicated
flavor structure that the one appearing in minimal gauge mediation
or supergravity schemes.
Observe that the contributions of 
the charged Higgs boson
and top-squark loops to  $b \rightarrow s \gamma$,
should be computed including the 
effect of the bottom mass corrections 
$\Delta(m_b)$ in the definition of the bottom Yukawa coupling $h_b$.
A next--to--leading--order 
SUSY QCD calculation of B($b \rightarrow s \gamma$) 
can be found in Ref.~\cite{CGGD}.

\section{Conclusions}

We have presented a study of some of the MSSM Higgs boson
signatures of relevance to the Tevatron collider. 
We first analyzed the
possibility of finding the lightest or heaviest CP--even Higgs
bosons in the $W\phi$ channel. 
Quite generally, for moderate and large $\tan \beta$,
 either the lightest or the heaviest CP--even Higgs boson has 
SM--like couplings to the vector bosons.   Therefore, 
most of the $\tan \beta-m_A$ plane is covered in this channel 
by producing the corresponding Higgs boson
with mass below 130 GeV, provided there is sufficient integrated luminosity, 
of the order of 30 fb$^{-1}$.
However for $m_A = m_{\phi^{SM}} \simeq 110-130$ GeV and large $\tan \beta$,
neither $h$ nor $H$ has
 SM--like couplings to the vector gauge bosons
and the coverage is decreased.
This problem is more pronounced for large values of the stop masses 
and mixing parameters.
For smaller values of $\tan \beta \simeq 5$, this problematic 
region extends to smaller values of $m_A$.
 In these cases,  very large luminosity,
above 30 fb$^{-1}$, will be needed or
the contribution of other production processes
will be necessary to assure full coverage.
To cover the region of $\tan\beta>10$ and $m_A\simeq m_{\phi^{SM}}$,
we have combined the two CP--even Higgs boson signals when their
masses are close to each other.
Another possibility may be to explore the $hA$ and $HA$ channels,
which will suffer from similar
suppression factors in the production cross sections ($\sin^2(\beta-\alpha) 
\simeq \cos^2(\beta-\alpha) \simeq {\cal{O}}(0.5)$), but may be   
combined with the $W\phi$ production process.

Furthermore, the branching ratio for the decay of the CP--even
Higgs bosons into bottom quarks
can be very different from the standard model one.
In particular, this takes place for moderate and  large $\tan\beta$,
when the off--diagonal elements of the Higgs boson mass matrix
can be strongly modified by radiative corrections induced
mainly by top squark--loops. 
We derived an approximate expression in terms of the MSSM
parameters to clarify when this occurs, 
and provided examples 
when the decays into bottom quarks are suppressed 
in a large region of parameter space,
thereby negatively affecting the Tevatron reach.

We have also emphasized that, due to supersymmetry breaking
effects, the values of the bottom and $\tau$ Yukawa couplings to the
CP--even Higgs bosons may be different from the ones computed
including only standard QCD corrections. Indeed, 
non--decoupling effects induced by supersymmetry breaking can
become particularly important for large $\tan\beta$, 
leading to modifications in the Higgs boson discovery and
exclusion potentials at the Tevatron. For instance,
if the bottom mass 
corrections $\Delta(m_b)$
are of order one, the decay of the standard
model--like Higgs boson into $\tau^+\tau^-$ may be enhanced, while
decays to $b\bar b$ are suppressed.
These supersymmetry breaking effects on the bottom and the 
$\tau$ Yukawa couplings 
can also  have an impact in the phenomenology of the 
charged Higgs boson. In particular, they can be relevant for
determining
the Tevatron limits on top decays into charged
Higgs bosons at large $\tan\beta$~\cite{sola,SOLAlast}.

The bottom Yukawa coupling corrections are particularly important for the
$b\bar b\phi$ process,  
because the production cross section
is proportional to the square of the bottom Yukawa coupling. We 
performed a phenomenological study to investigate the relevance of these
corrections.  Even with large luminosity factors, of the order
of 30 fb$^{-1}$, 
and negative bottom mass corrections, which enhance the production
rate,  the Tevatron can discover a CP--odd
Higgs boson 
(together with  a CP--even Higgs boson with mass and couplings similar to it)
only if its mass is not  larger than about 200--300 GeV.
The reach is only efficient
for moderate or large values of $\tan\beta$. 
Such values for the CP--odd Higgs boson mass 
give positive contributions to $B(b \rightarrow s \gamma)$,
and the discovery of such a Higgs boson would 
constrain the masses
and mixing angles of the top squarks unless  a non--trivial mixing
between the second and third generation down squarks is present.

The computation of the Higgs boson
mass matrix elements considered in this article~\cite{CAESQUWA}
is still affected by theoretical uncertainties, most notably,
those associated with the two--loop, finite, threshold corrections to the
effective quartic couplings of the Higgs potential. Recently, a partial,
diagrammatic, two--loop computation of the Higgs mass 
has been performed~\cite{HEHOWE}. 
In the limit of large $m_A$, these
additional contributions lead to a slight modification of
the dependence of the lightest CP--even Higgs boson mass
on the stop mixing parameters.
For instance, although the upper bound on the lightest CP--even
Higgs boson mass for squark masses of approximately 1 TeV
is approximately the same as the one obtained
to next--to--leading--order accuracy (as done in this work),
the upper bound on 
the Higgs boson mass is reached for values
of $|\tilde{a}| \simeq 2$ instead of $|\tilde{a}| = \sqrt{6}$,
and has a weak dependence on the sign of $\tilde a$.
A diagrammatic computation of the  two--loop corrections 
induced by the top Yukawa coupling, which are  included at the
leading--logarithmic level 
in our computation, is, however, still lacking.~\cite{Zhang}

Summarizing, at present, the $W\phi$ channel with the $W$ decaying 
leptonically 
and the Higgs boson decaying into $b$ quarks remains the golden mode
to test the MSSM Higgs sector at the Tevatron. 
The other channel we have  analyzed, $b\bar b \phi$ production 
with the subsequent decay of $\phi$ into $b$ quarks and $\tau$ leptons,
proves to be very useful to cover regions of large $\tan \beta$ and small 
to moderate $m_A$ up to about
250 GeV. However, the reach in these channels requires a large
total integrated luminosity.
Because of this,
other production processes and Higgs boson decay modes need to be carefully 
investigated
if we want to fully probe the MSSM Higgs sector at the Tevatron.
We have identified regions of parameter space where the Higgs decay 
into $b \bar b$ is strongly suppressed.
In these regions,
other search techniques will have to be considered, due to the
presence of enhanced decays to $W^*W^*$, $gg$
and $c\bar c$ final states.   
Clearly, there is a motivation
to study these final states, and in particular the $W^*W^*$ one,
 even for lighter Higgs bosons for which the SM Higgs boson decay rate is 
strongly suppressed.
In addition other production processes like the associated 
production of $hA$ or $HA$ 
with the subsequent decays into $b$ quarks and $\tau$ leptons
may also be useful.

A careful study of all different possibilities, which may be relevant in 
different regions of parameter space, and the combination of channels may
allow a full coverage of the MSSM parameter space with luminosities achievable
at the Tevatron. 
If that is the case, the Tevatron can discover a light 
Higgs boson which might be beyond
the presently expected LEP2 reach for 
generic values of the supersymmetric mass 
parameters.   
Most importantly, the detection of one or more Higgs bosons at the Tevatron 
will give very valuable information about the Higgs and stop sectors of 
the MSSM.

In the final stages of this work, two preprints appeared on
related topics.  One addressed the issue of the $W\phi$ reach
of the Tevatron collider \cite{baer_harris_tata}.  
The other commented on the possible effects of large $\tan\beta$
corrections to $b\bar b\phi$ production at hadron colliders
\cite{nafta}.
In the special cases when the analyses are comparable 
we tend to agree with their results, although the authors of
Ref.~\cite{baer_harris_tata} do not see any visible dependence
on the sign of $\mu \times \tilde{a}$.
The present  work goes beyond those studies by providing 
a detailed numerical and theoretical
analysis of the dependence of the Tevatron discovery potential on the
MSSM parameter space.

\acknowledgments

CEW thanks the hospitality of the
theory groups at Fermilab and at the University
of Buenos Aires, where part of this work was completed.
MC and CEW are grateful to the Rutherford Laboratory, as
is SM to the Aspen Center for Physics.
We also acknowledge discussions with K. Matchev and T. Tait.
The research of MC is supported by the Fermi National Accelerator
Laboratory, which is operated by the Universities Research Association,
Inc., under contract no. DE-AC02-76CHO300.  The work of SM is
supported in part by the U.S. Dept. of Energy, High Energy Physics
Division, under contract W-31-109-ENG-38.

%
%
\begin{center}
\begin{figure}
\psfig{figure=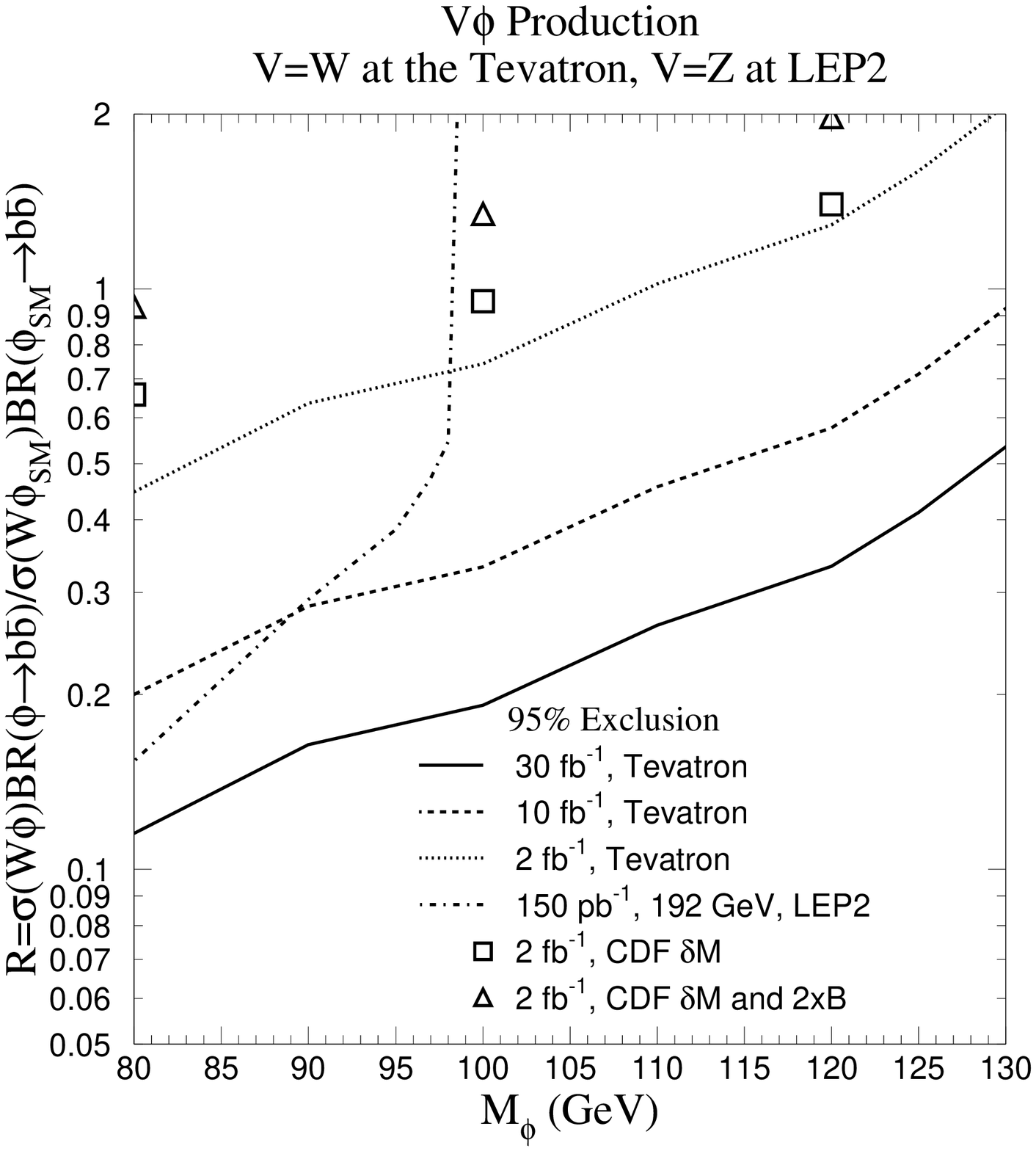,width=14cm}
\caption[]
{95\% C.L. bound on
$R=\frac{\sigma(p \bar p \to W \phi)}{\sigma(p \bar p \to W\phi^{SM})}
 \frac{B(\phi\to b\bar b)}{B(\phi^{SM}\to b\bar b)}$
as a function of
the Higgs boson mass
for the Tevatron and LEP2.}
\label{fig:95cl}
\end{figure}
\begin{figure}
\psfig{figure=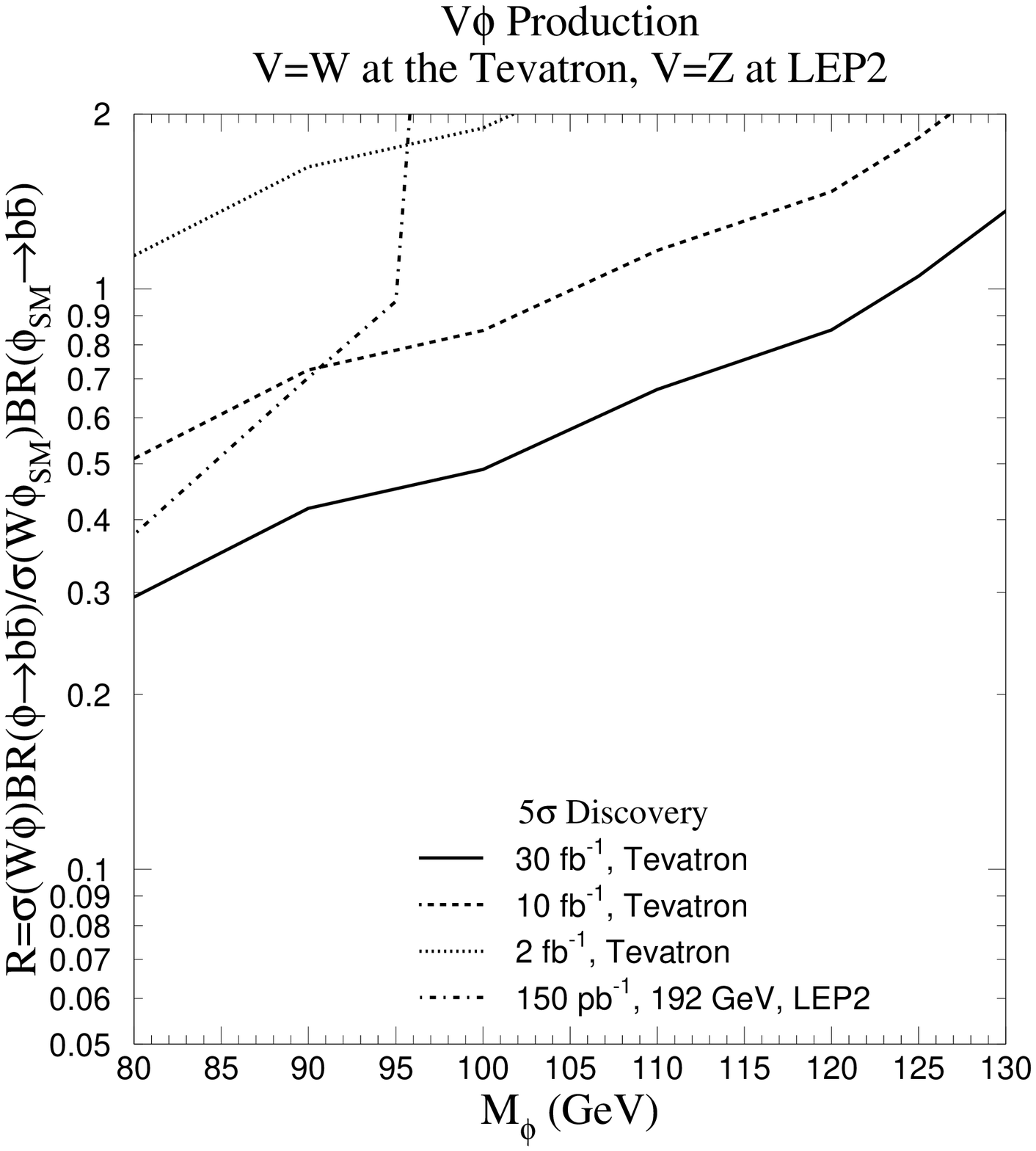,width=14cm}
\caption[]
{Same as Fig.~1 except for $5\sigma$ discovery.}
\label{fig:5sig}
\end{figure}
\begin{figure}
\psfig{figure=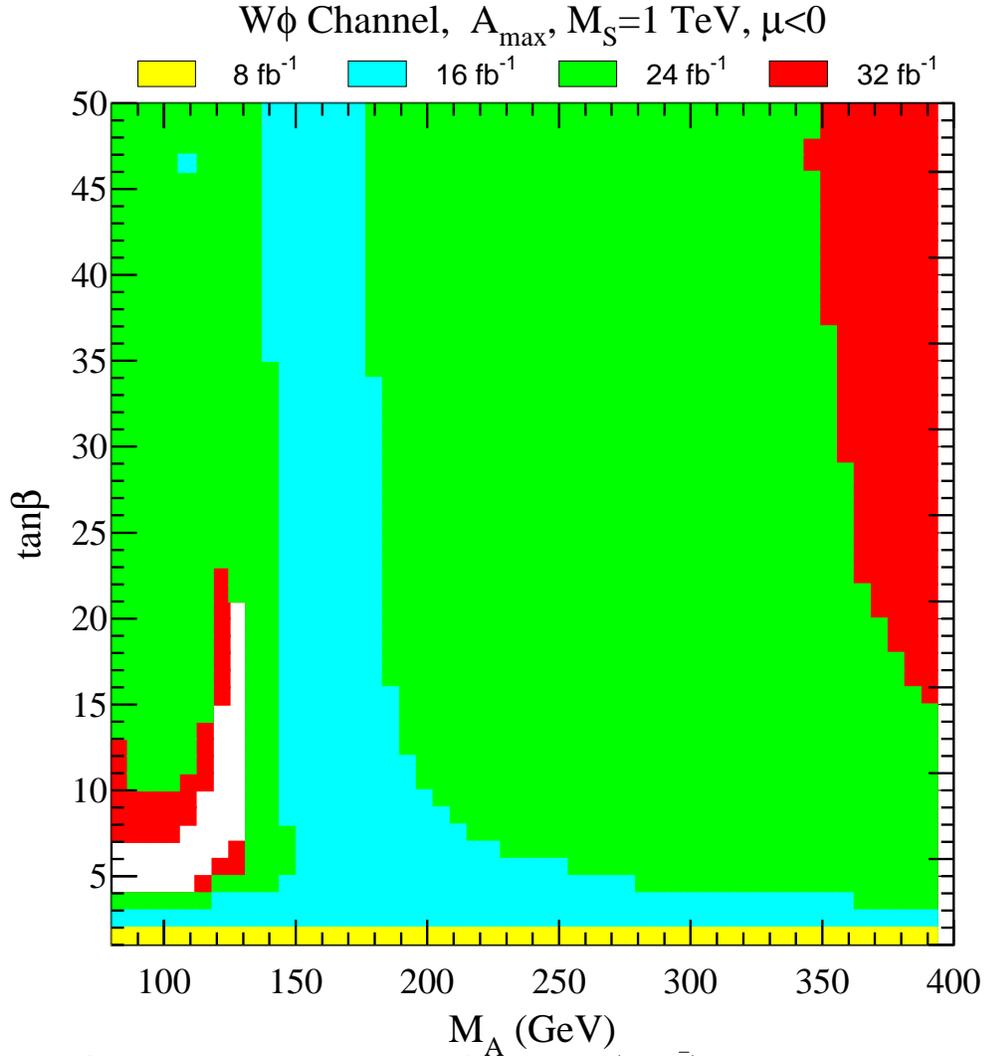,width=14cm}
\caption[]
{$5\sigma$ discovery contours for the
$W\phi(\to b\bar b)$ mode at the Tevatron in
the MSSM for maximal mixing, $\mu<0$, and $M_S$=1 TeV.
Different shadings correspond to different
integrated luminosities.}  
\label{fig:max_m}
\end{figure}
\begin{figure}
\psfig{figure=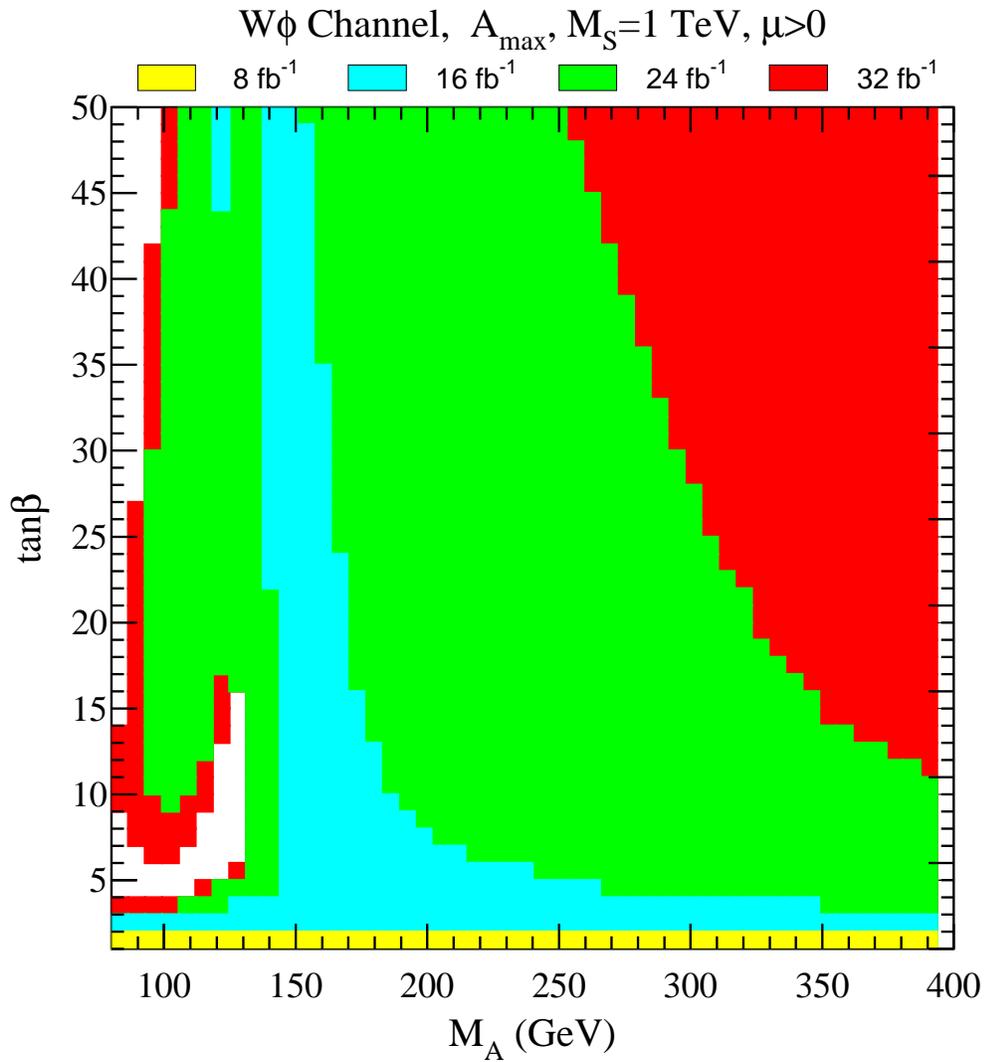,width=14cm}
\caption[]
{Same as Fig.~3 but with $\mu>0$.}
\label{fig:max_p}
\end{figure}
\begin{figure}
\psfig{figure=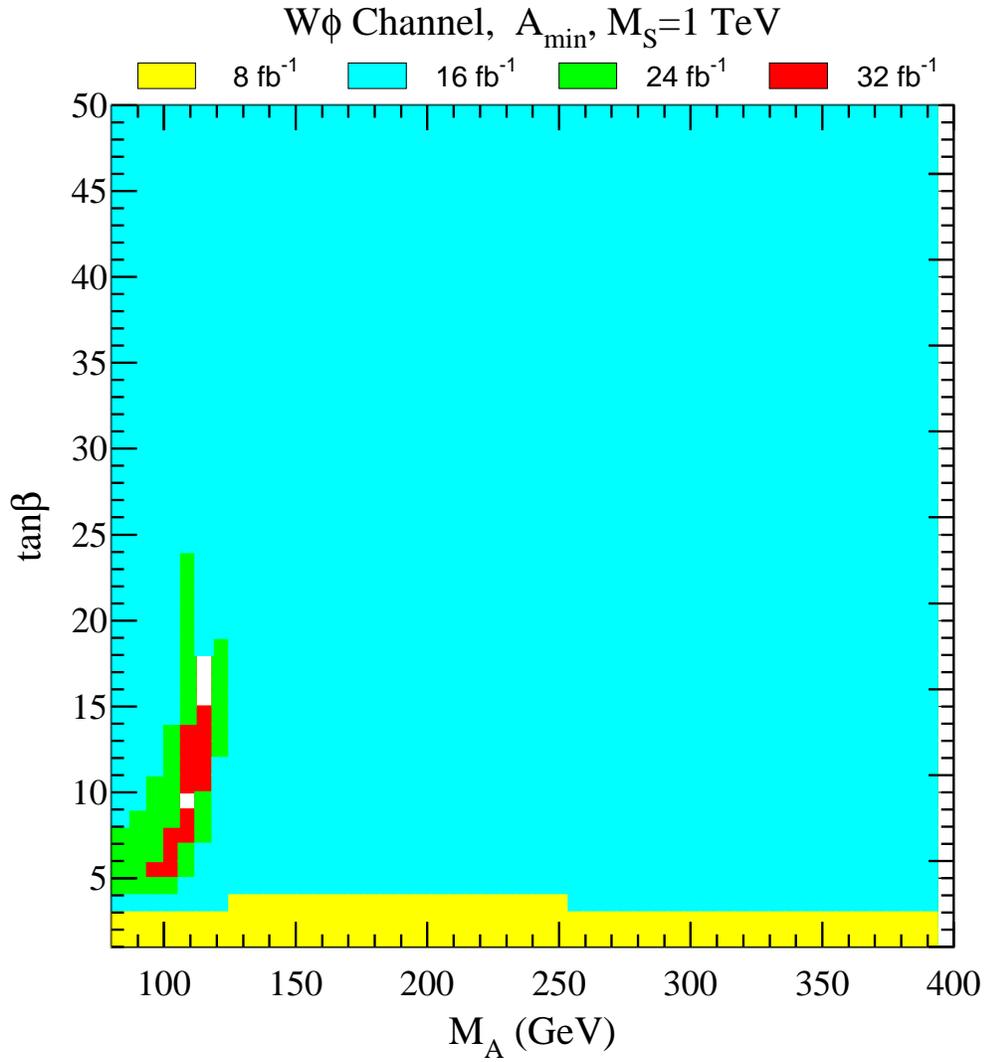,width=14cm}
\caption[]
{Same as Fig.~3 but for minimal mixing.}
\label{fig:min_p}
\end{figure}
\begin{figure}
\psfig{figure=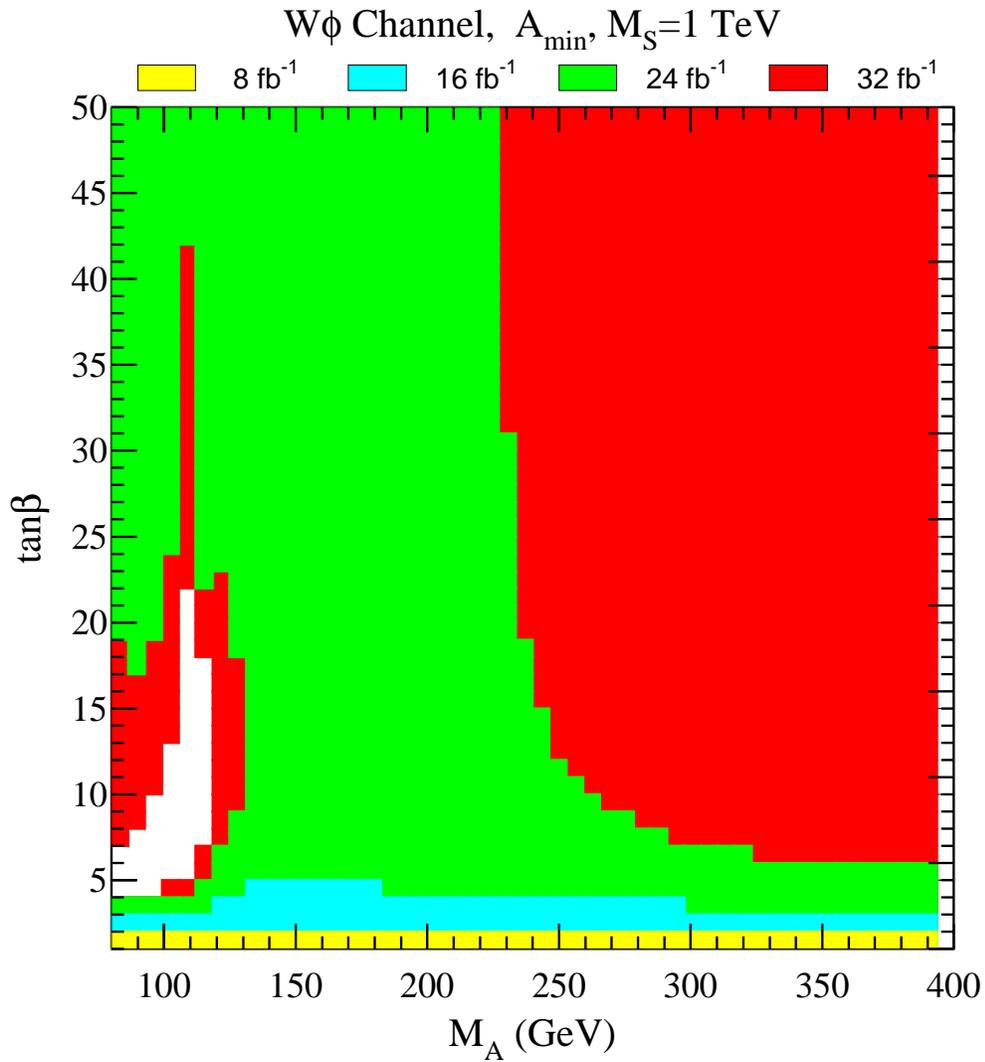,width=14cm}
\caption[]
{Same as Fig.~5 but using the present CDF mass resolution.}
\label{fig:min_p_kuhl}
\end{figure}
\begin{figure}
\psfig{figure=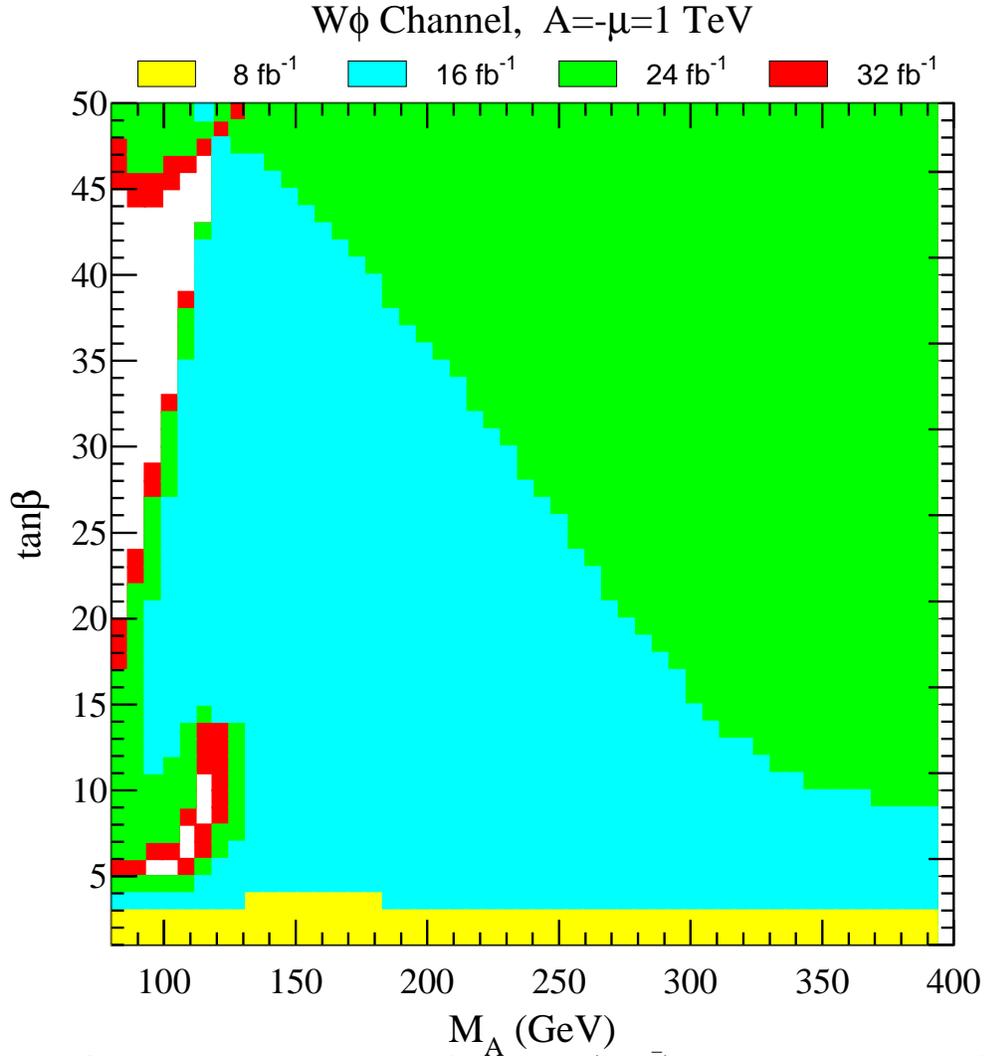,width=14cm}
\caption[]
{$5\sigma$ discovery contours for
the $W\phi(\to b\bar b)$ mode at the Tevatron 
for $A=-\mu=$ 1 TeV and $M_S=1$ TeV.}
\label{fig:amu1}
\end{figure}
\begin{figure}
\psfig{figure=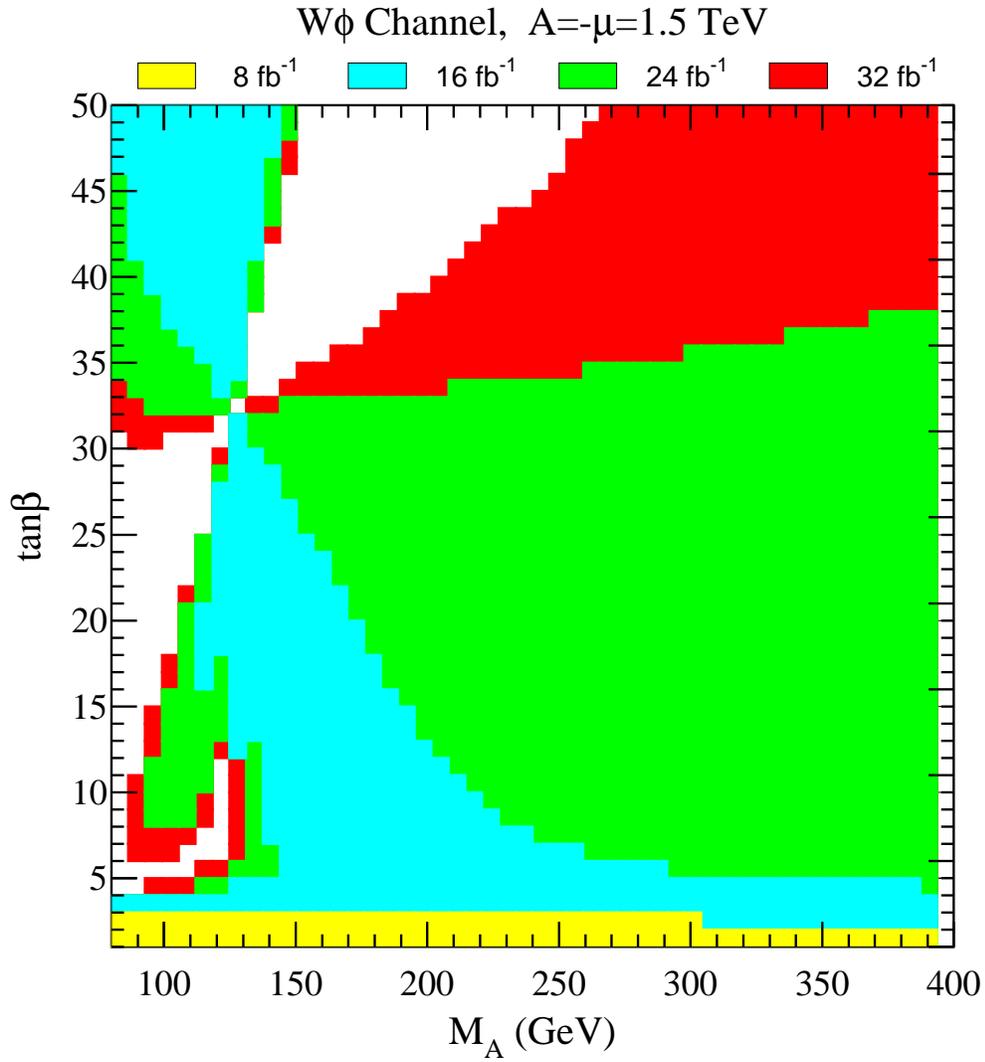,width=14cm}
\caption[]
{Same as Fig.~\ref{fig:amu1} but
for $A=-\mu=$ 1.5 TeV.}
\label{fig:amu15}
\end{figure}
\begin{figure}
\psfig{figure=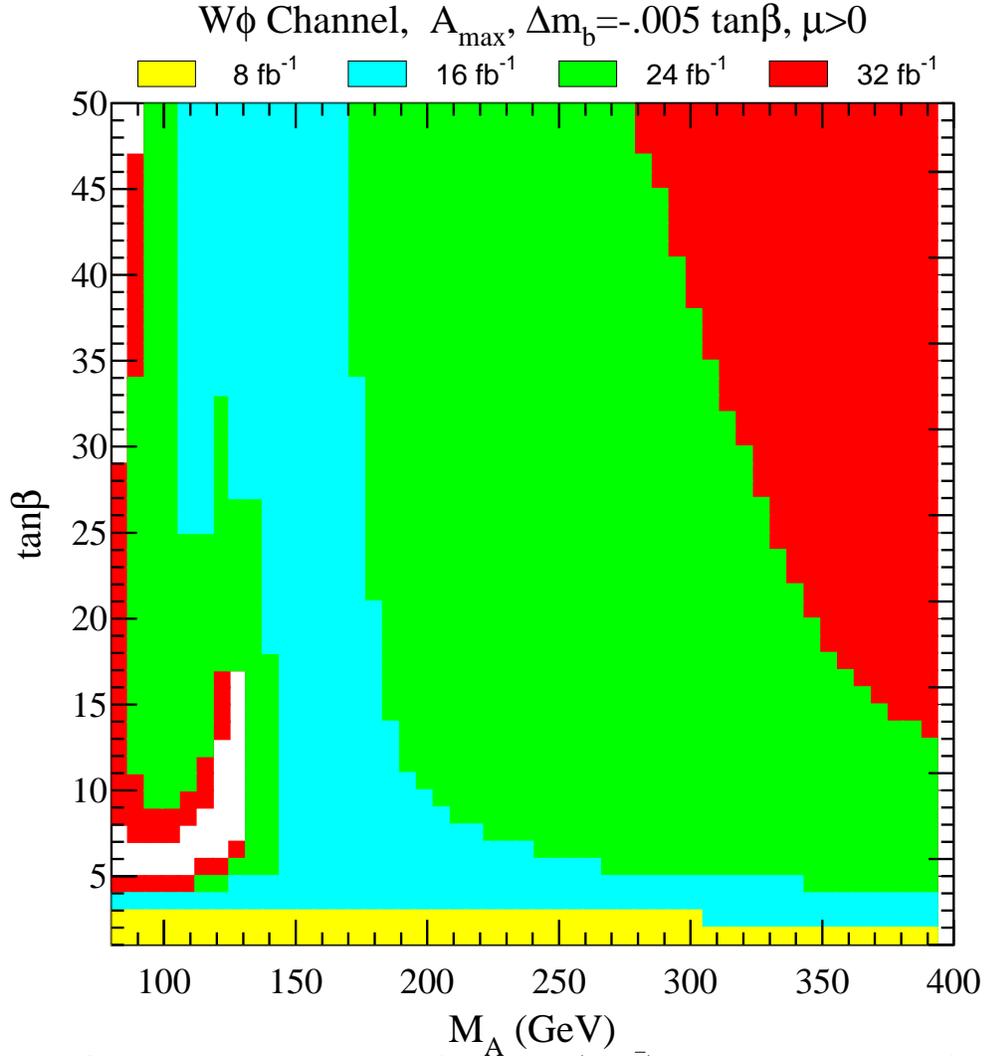,width=14cm}
\caption[]
{$5\sigma$ discovery contours for the
$W\phi(\to b\bar b)$ mode at the Tevatron 
for maximal mixing and $\mu>0$ after including
large $\tan\beta$ corrections: $\Delta(m_b)=-.005\tan\beta$.
Different shadings correspond to different
integrated luminosities.}  
\label{fig:max_p_m005}
\end{figure}
\begin{figure}
\psfig{figure=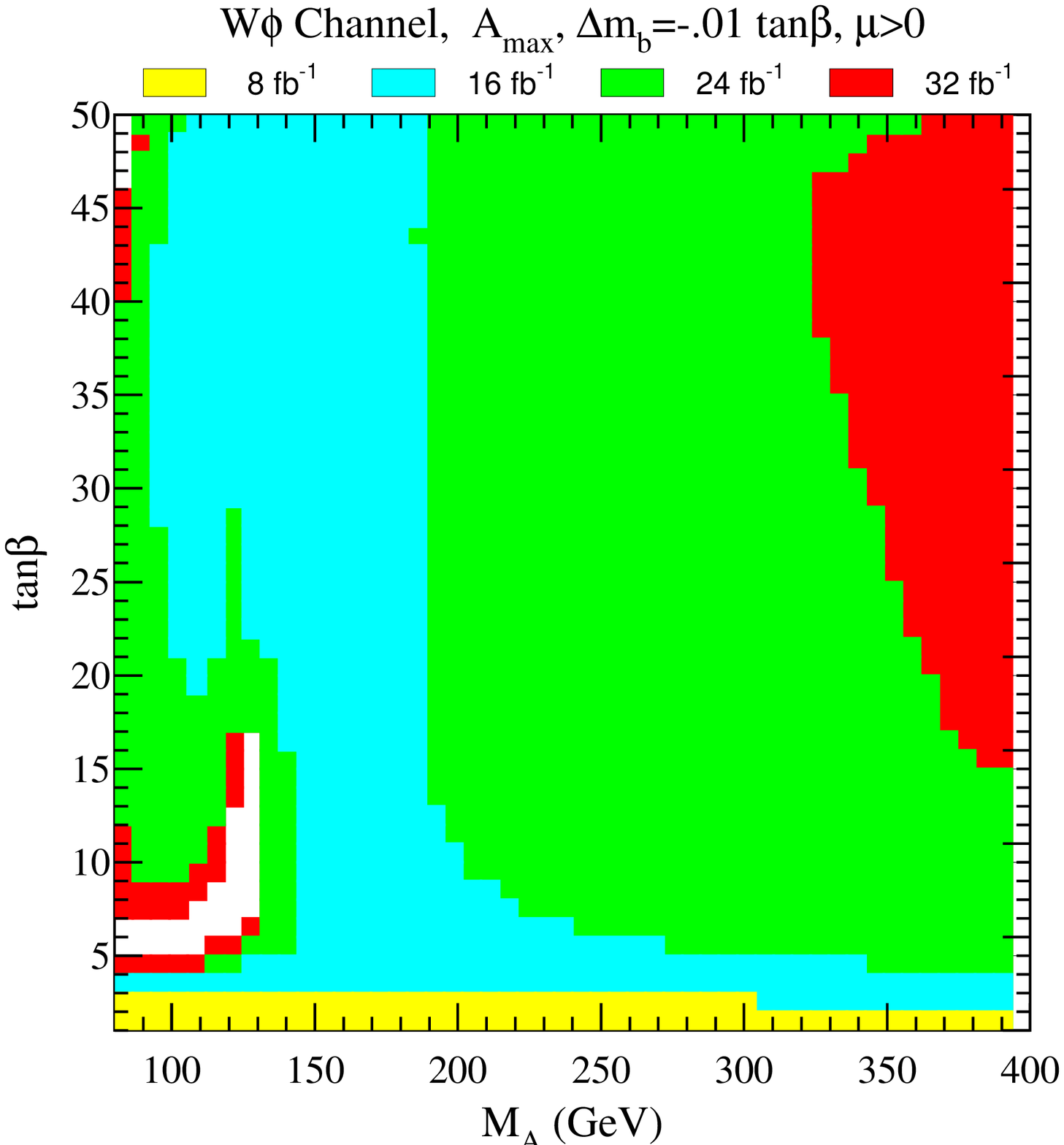,width=14cm}
\caption[]
{Same as Fig.~\ref{fig:max_p_m005} but for
$\Delta(m_b)=-.01\tan\beta$.}
\label{fig:max_p_m01}
\end{figure}
\begin{figure}
\psfig{figure=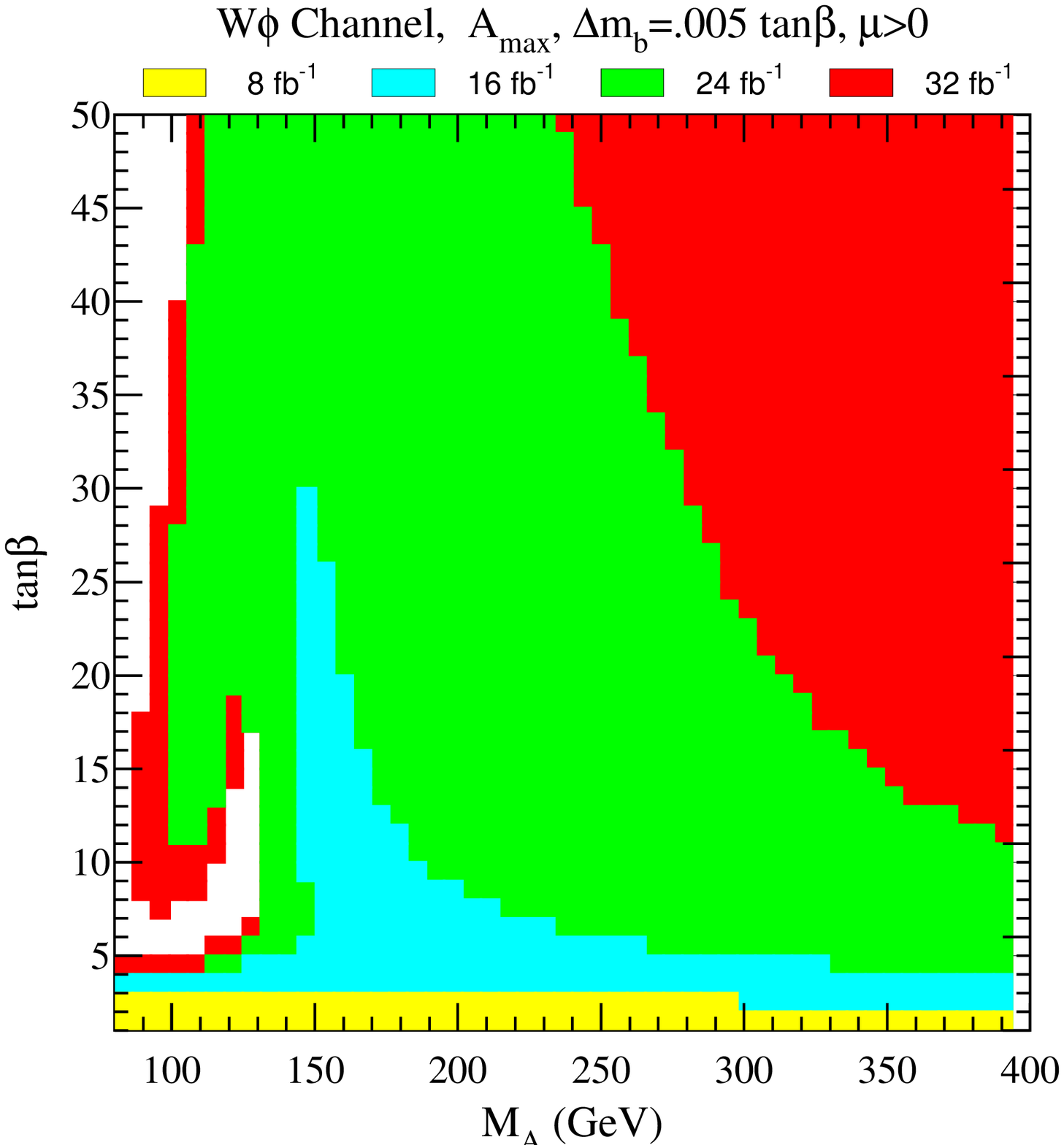,width=14cm}
\caption[]
{Same as Fig.~\ref{fig:max_p_m005} but for
$\Delta(m_b)=.005\tan\beta$.}
\label{fig:max_p_p005}
\end{figure}
\begin{figure}
\psfig{figure=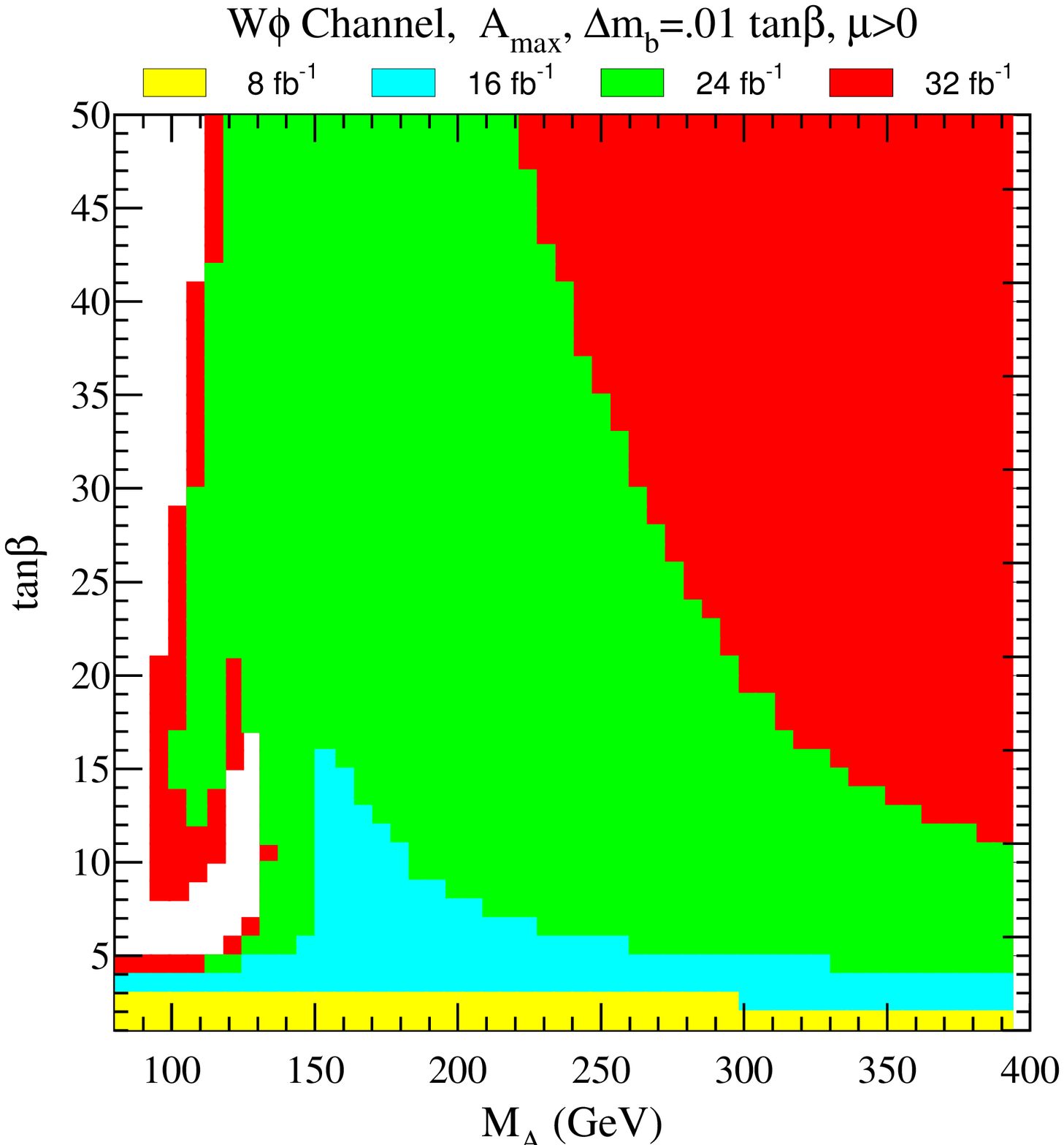,width=14cm}
\caption[]
{Same as Fig.~\ref{fig:max_p_m005} but for
$\Delta(m_b)=.01\tan\beta$.}
\label{fig:max_p_p01}
\end{figure}
%
\begin{figure}
\psfig{figure=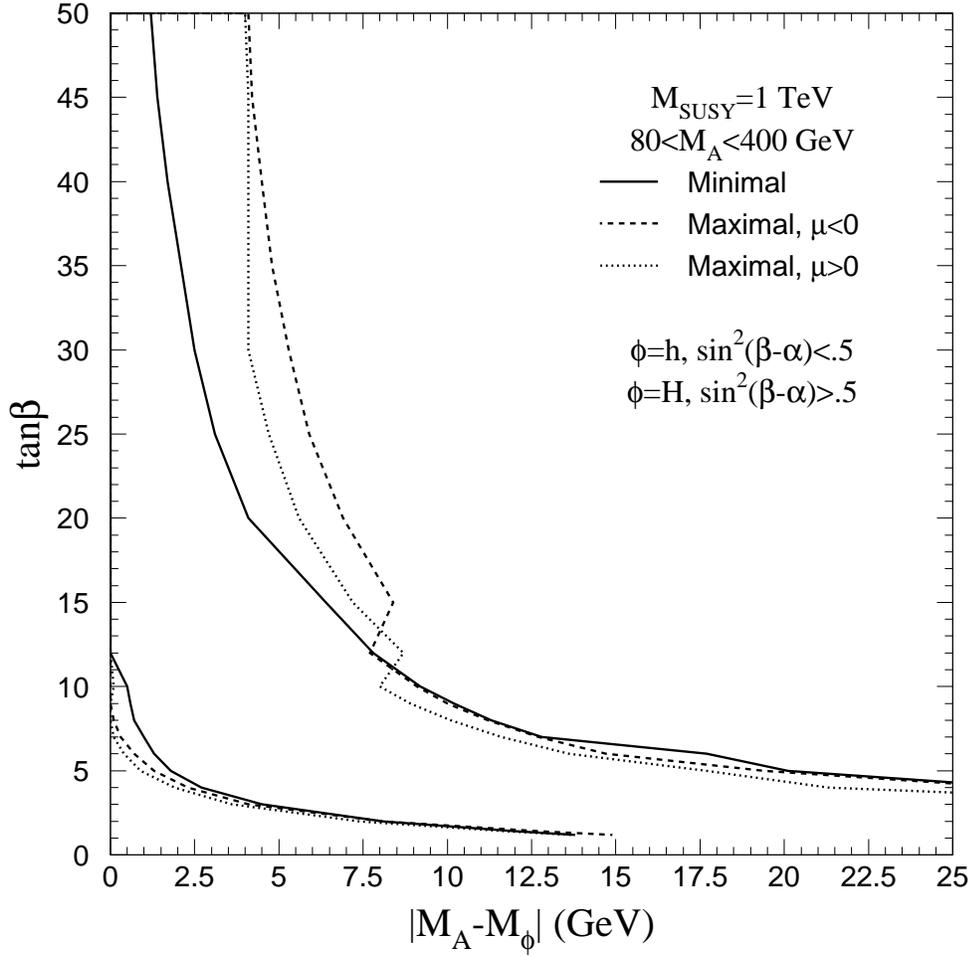,width=14cm}
\caption[]{Mass difference between the CP--odd 
Higgs boson and the CP--even Higgs boson (with properties
similar to the CP--odd one) as a function 
of $\tan\beta$ for several
MSSM parameter choices.}
\label{fig:dmass}
\end{figure}
\begin{figure}
\psfig{figure=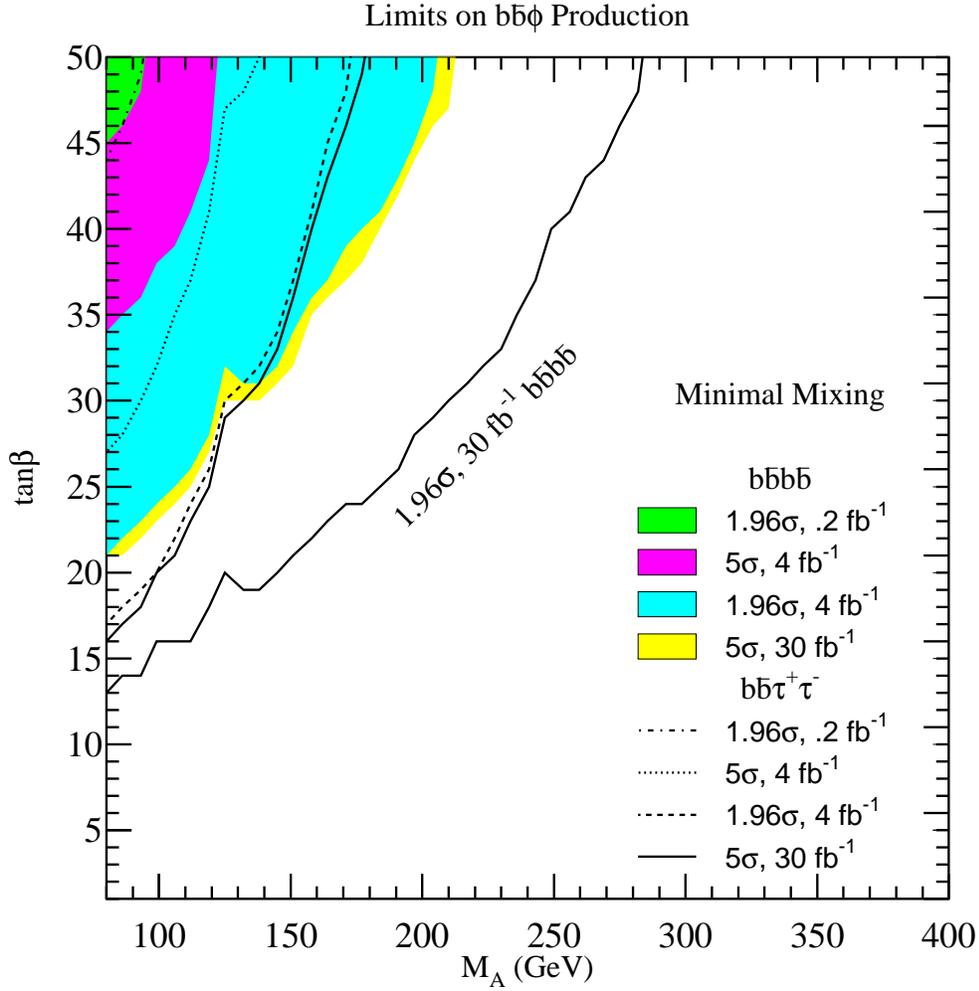,width=14cm}
\caption{95\% C.L. exclusion and $5\sigma$ discovery contours for 
$b\bar b\phi$ production at the Tevatron in
the MSSM for minimal mixing.  The shaded areas
correspond to different exclusion and discovery
contours at different integrated luminosities
for the $b\bar b b\bar b$ final state.  The different
lines show the same for the $b\bar b\tau^+\tau^-$ final state,
except that
the lower solid line shows the 95\% C.L. exclusion contour
for the $b\bar b b\bar b$ final state and 30 fb$^{-1}$.}
\label{fig:bbbb}
\end{figure}
\begin{figure}
\psfig{figure=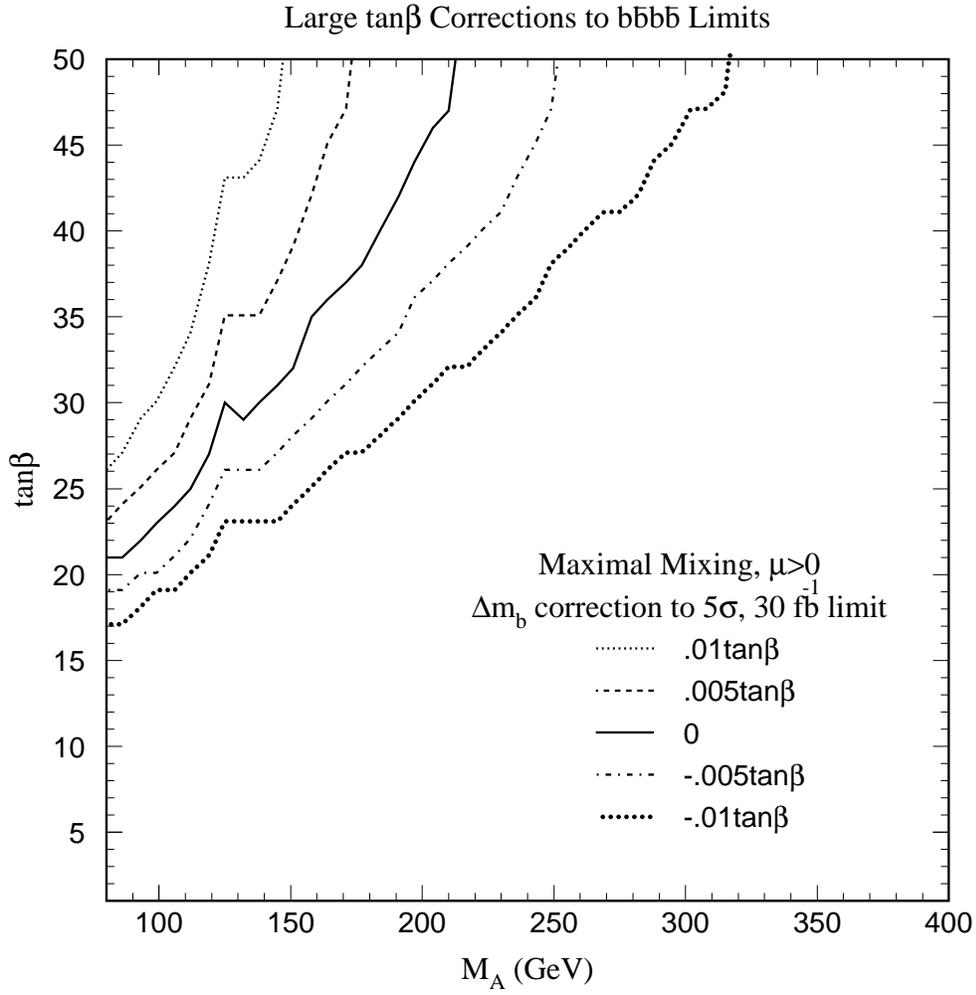,width=14cm}
\caption{$5\sigma$ discovery contours for the
$b\bar b\phi(\to b\bar b)$ mode and 30 fb$^{-1}$
with maximal mixing, $\mu>0$,
and $\Delta(m_b)=(0,\pm .005,\pm .01)\tan\beta$.}
\label{fig:bbbb1}
\end{figure}
\end{center}
\end{document}